\documentclass{aa}
\usepackage[varg]{txfonts}

\usepackage{graphicx} 
\usepackage{csquotes}
\usepackage{bm}
\usepackage{amsmath}
\usepackage{subcaption}
\usepackage{booktabs}
\usepackage{grffile}
\usepackage{overpic}
\usepackage{float}
\usepackage{enumitem}
\usepackage{amsfonts}
\usepackage{easy-todo}
\usepackage{listings}
\usepackage{lineno}
\usepackage[citecolor=blue]{hyperref}

\hypersetup{
    colorlinks=true,
    linkcolor=blue,
    filecolor=magenta,      
    urlcolor=cyan,
}
\usepackage{color}

\newcommand{\ten}[1]{{\bm #1}}
\renewcommand{\vec}[1]{{\bm #1}}

\newcommand{\mycaption}[2]{\caption[#1]{\emph{#1} #2}}

\newcommand{\dm}[2]{#1}{} % #1:new text in red; #2 old text removed

\newcommand{\rs}[2]{#1}{} % #1:new text in red; #2 old text removed
\allowdisplaybreaks

\newcommand\thefontsize{The current font size is: \f@size pt}
\begin{document}

\bibpunct{(}{)}{;}{a}{}{,} % to follow the A&A style

% Set image location
\graphicspath{{images/}}

% Avoid a title too similar to the last paper:
\title{Kelvin-Helmholtz instability and collapse of a twisted magnetic null point with anisotropic viscosity}
\titlerunning{Kelvin-Helmholtz instability and collapse of a twisted magnetic null point with anisotropic viscosity}
\authorrunning{Quinn et al.}

%%% Authors
\author{
  James Quinn\inst{1}\thanks{\href{https://orcid.org/0000-0002-0268-7032}{orcid.org/0000-0002-0268-7032}} \and
  David MacTaggart\inst{2}\thanks{\href{https://orcid.org/0000-0003-2297-9312}{orcid.org/0000-0003-2297-9312}} \and
  Radostin Simitev\inst{2}\thanks{\href{https://orcid.org/0000-0002-2207-5789}{orcid.org/0000-0002-2207-5789}}
}
\institute{
  Research Software Development Group, University College London, Gower Street, London WC1E 6BT, UK email:\href{mailto://jamiejquinn@jamiejquinn.com}{jamiejquinn@jamiejquinn.com} \and
  School of Mathematics and Statistics, University of Glasgow, Glasgow G12 8QQ, UK 
}

\date{\rs{Version dated \today}{TODO / TODO}}

\abstract{
  Magnetic null points are associated with high-energy coronal phenomena such as solar flares and are often sites of reconnection and particle acceleration. Dynamic twisting of a magnetic null point can generate a Kelvin-Helmholtz instability (KHI) within its \rs{fan plane}{fan-plane} and, under continued twisting, can instigate spine-fan reconnection and an associated collapse of the null \rs{point}{}.
}{
   This article aims to compare the effects of isotropic and anisotropic viscosity in simulations of the KHI and collapse in a dynamically twisted magnetic null point.
}{
  Simulations were performed using the 3D magnetohydrodynamics code Lare3d with a custom anisotropic viscosity module. A pair of high resolution simulations \dm{was}{were} performed, one using isotropic viscosity and another using anisotropic viscosity, keeping all other factors identical, and the results analysed in detail. A further parameter study was performed over a range of values for viscosity and resistivity. 
}{
  Both viscosity models permit the growth of the Kelvin-Helmholtz instability and the eventual collapse of the null \rs{point}{}. Over all studied parameters, anisotropic viscosity allows a faster growing instability, while isotropic viscosity damps the instability\rs{}{,} to the \rs{extent}{point} of stabilisation in some cases. Although the viscous heating associated with anisotropic viscosity \dm{is}{was} generally smaller, the \rs{ohmic}{Ohmic} heating dominates and \dm{is}{was} enhanced by the current sheets generated by the instability, leading to a greater overall heating rate when using anisotropic viscosity. The collapse of the null \rs{point}{} \dm{occurs}{occurred} significantly sooner when anisotropic viscosity \dm{is}{was} employed.
}{}

\keywords{viscosity - magnetohydrodynamics - Kelvin-Helmholtz instability - magnetic null point - anisotropic viscosity - Sun: corona - Sun: magnetic fields}

\maketitle

\section{Introduction}

This paper presents the results of a series of numerical experiments intended to develop an understanding of the effect of anisotropic viscosity on the Kelvin-Helmholtz instability (KHI) in the fan plane of a magnetic null point, reproducing and extending the work of~\citet{wyperKelvinHelmholtzInstabilityCurrentvortex2013}. We continue to stress the null point beyond the time investigated in~\citet{wyperKelvinHelmholtzInstabilityCurrentvortex2013}, \rs{which allows}{allowing} us to also study the effect of anisotropic viscosity on the spontaneous collapse of the null point. The experiments take the form of \rs{a}{the} dynamic twisting of an initially static magnetic null point at the footpoints of its spine, resulting in a current-vortex sheet which forms in the fan plane and can be unstable to the KHI, given appropriate parameter choices. All experiments are carried out using both isotropic and anisotropic viscosity over a range of parameter choices. Anisotropic viscosity is modeled following \cite{mactaggartBraginskiiMagnetohydrodynamicsArbitrary2017}. Continued driving after the \rs{moment}{point} at which the KHI occurs causes the null \rs{point}{} to spontaneously undergo spine-fan reconnection and collapse.

The KHI has been \rs{well-studied}{well studied} in \rs{the magnetohydrodynamic (MHD) context}{MHD} and can be found in a number of coronal \rs{situations}{contexts}, both in numerical simulations~\citep{howsonEffectsResistivityViscosity2017,wyperKelvinHelmholtzInstabilityCurrentvortex2013} and in observations~\citep{foullonMAGNETICKELVINHELMHOLTZINSTABILITY2011,yangObservationKelvinHelmholtz2018}. See~\citet{faganelloMagnetizedKelvinHelmholtz2017} for a recent review of the KHI in MHD and~\citet{chandrasekhar1961} for a classical treatment. 

In general, the effect of a magnetic field is stabilising; when the wavevector of a perturbation in a velocity shear layer is parallel or at an oblique angle to a magnetic field, magnetic tension acts to stabilise the KHI~\citep{chandrasekhar1961,ryuMagnetohydrodynamicKelvinHelmholtzInstability2000}. Otherwise, the KHI acts as an interchange instability and the magnetic field does not affect its linear stability~\citep{chandrasekhar1961}.

In a current-vortex sheet, where a velocity shear coincides with a magnetic shear, the balance of shear layer strength and \rs{thickness}{thickness} dictates if the KHI, \rs{the}{} tearing instability, or some mixture \rs{of them}{}, is excited. Generally, when the magnetic shear is strong compared to the velocity shear, the KHI is suppressed and the tearing instability grows~\citep{einaudiResistiveInstabilitiesFlowing1986}. This can be somewhat modified by the inclusion of viscosity~\citep{einaudiResistiveInstabilitiesFlowing1989}. The nonlinear development of the KHI is known to enhance reconnection by the local distortion of magnetic field lines, the generation of current sheets~\citep{minEffectsMagneticReconnection1997} and by generating local turbulence in conjunction with the tearing instability~\citep{kowalKelvinHelmholtzTearingInstability2020}.

The effect of (anisotropic) viscosity on the stability of a current-vortex sheet is to suppress the growth of the KHI, although viscosity is found to enhance the linear growth of the tearing instability, when the KHI is stabilised by a strong magnetic field~\citep{einaudiResistiveInstabilitiesFlowing1989}. A number of studies suggest isotropic viscosity can also slow and even suppress the KHI~\citep{howsonEffectsResistivityViscosity2017,roedigerViscousKelvinHelmholtzInstabilities2013a,wyperKelvinHelmholtzInstabilityCurrentvortex2013}.

Magnetic null points\dm{}{, locations in a magnetic field where the field strength goes to zero,} are an abundant feature in the topologically complex coronal magnetic field~\citep{edwardsNullPointDistribution2015}. Given they are sites coinciding with changes in topology, they are strongly associated with reconnection processes~\citep{yangImagingSpectralStudy2020,sunHOTSPINELOOPS2013}. Additionally they are inferred to participate in a number of high-energy phenomena, such as in the generation of flare ribbons in compact solar flares~\citep{massonNATUREFLARERIBBONS2009,pontinWhyAreFlare2016a} and the production of jets~\citep{moreno-insertisPLASMAJETSERUPTIONS2013} and of coronal mass ejections~\citep{barnesRelationshipCoronalMagnetic2007,zouContinuousNullPointMagnetic2020}, particularly through \dm{their key involvement in the breakout model of solar eruptions }{ their necessary involvement in the breakout model of eruptive solar flares~}\citep{AntiochosCME199,macleanTopologicalAnalysisMagnetic2005}.

A much-studied form of reconnection in 3D null points is spine-fan reconnection, where a strong current sheet forms in the vicinity of the null point and enables efficient reconnection between the magnetic field making up the spine and fan plane, collapsing the field around the null in the process~\citep{thurgoodImplosiveCollapseMagnetic2018}. The collapse of a null point has the potential to develop into a form of oscillatory reconnection~\citep{McLauglinOscRec2012,thurgoodThreedimensionalOscillatoryMagnetic2017}.

The layout of this paper is as follows. The numerical setup of the simulations is presented in section~\ref{sec:khi_numerical_setup}, including a description of the model of the linear null point and footpoint driver. The methods of calculating stability measures, shear layer properties and reconnection rate are described in section~\ref{sec:khi_analysis}. In the first part of section~\ref{sec:khi_results} the results of a high-resolution pair of simulations \rs{}{are presented} for a single choice of viscosity and resistivity parameters \rs{are presented}{} and the \rs{effects}{effect} of the viscosity model \rs{used are}{} compared. In the second part, the results of a parameter study are \rs{described}{presented}, generalising the high-resolution results. The chapter concludes with a discussion of findings in section~\ref{sec:khi_discussion} and conclusions in section~\ref{sec:khi_conclusions}.

\section{Model and numerical setup}
\label{sec:khi_numerical_setup}

\subsection{Governing equations}

\rs{We consider}{The governing equations are} the non-dimensionalised visco-resistive MHD equations,
\begin{gather}
\label{eq:mhda}
\frac{D\rho}{Dt} = - \rho \vec{\nabla} \cdot \vec{u},\\
\rho\frac{D\vec{u}}{Dt} = -\vec{\nabla} p + \vec{\jmath} \times \vec{B} + \vec{\nabla} \cdot \ten{\sigma},\\
\frac{D\vec{B}}{Dt} = (\vec{B} \cdot \vec{\nabla})\vec{u} - (\vec{\nabla} \cdot \vec{u})\vec{B} + \eta \nabla^2 \vec{B},\\
\rho\frac{D\varepsilon}{Dt} = -p \vec{\nabla} \cdot \vec{u} + {Q}_{\nu} + {Q}_{\eta},
\label{eq:energy}
\end{gather}
\rs{posed in a rectangular domain $\Omega$,}{} where $\rho$ is the mass density, $\vec{u}$ is the plasma velocity, $p$ is the thermal pressure, $\vec{\jmath}$ is the current density, $\vec{B}$ is the magnetic field, $\ten{\sigma}$ is the viscous stress tensor, $\eta$ is the resistivity, equivalent to the inverse Lundquist number, and $\varepsilon$ is the specific energy density, given by the equation of state for an ideal gas,
\begin{equation}
\varepsilon = \frac{p}{\rho(\gamma - 1)},
\end{equation}
where the specific heat \rs{capacity}{} ratio is given by $\gamma = 5/3$. The terms ${Q}_{\nu} = \ten{\sigma} : \vec{\nabla}\vec{u}$ and ${Q}_{\eta} = \eta | \vec{\jmath} |^2$ are the viscous and \rs{ohmic}{Ohmic} heating contributions, respectively. $D/Dt = \partial/\partial t + (\vec{u}\cdot\vec{\nabla})$ is the material derivative.

The non-dimensionalisation scheme is identical to that used in the code Lare3d~\citep{arberStaggeredGridLagrangian2001}, where a typical magnetic field strength $B_0$, density $\rho_0$ and length scale $L_0$ are chosen and the other variables non-dimensionalised appropriately. Velocity and time are
non-dimensionalised using the Alfv\'en speed $u_A = B_0 / \sqrt{\rho_0
  \mu_0}$ and Alfv\'en crossing time $t_A = L_0/u_A$,
respectively. Temperature is non-dimensionalised via $T_0 = u_A^2
\bar{m} / k_B$, where $k_B$ is the Boltzmann constant and $\bar{m}$ is
the average mass of ions, here taken to be $\bar{m} = 1.2m_p$ (a mass
typical for the solar corona) where $m_p$ is the proton mass. Dimensional quantities can be recovered by multiplying the non-dimensional variables by their respective reference value (e.g. $\vec{B}_{\dim} = B_0 \vec{B}$). The reference values used here are $B_0 = 5 \times 10^{-3}$ T, $L_0 = 1$ Mm and $\rho_0 = 1.67 \times 10^{-12} \ \text{kg m}^{-3}$, giving reference values for the Alfv\'en speed $u_A = 3.45\ \text{Mm s}^{-1}$, Alfv\'en time $t_A = 0.29\ \text{s}$ and temperature $T_0 = 1.73 \times 10^{9}\ K$.

\subsection{Models of viscosity}

Both isotropic and anisotropic models of viscosity are used \rs{and compared}{} here. \rs{In the isotropic case, the Newtonian viscosity model is used}{Isotropic viscosity is modelled as Newtonian viscosity},
\begin{equation}
  \label{eq:isotropic_viscous_tensor}
  \ten{\sigma}_{\text{iso}} = \nu \ten{W},
\end{equation}
where $\nu$ is the viscous parameter (termed the viscosity throughout this paper) and $\ten{W}$ is the rate-of-strain tensor,
\begin{equation}
  \label{eq:rate_of_strain}
  \ten{W} = \nabla\vec{u} + (\nabla\vec{u})^T - \tfrac{2}{3}(\nabla \cdot \vec{u})\ten{I}.
\end{equation}
\rs{In the anisotropic case, the switching viscosity model of \citet{mactaggartBraginskiiMagnetohydrodynamicsArbitrary2017} is used}{Anisotropic viscosity is modelled using the switching model~\citep{mactaggartBraginskiiMagnetohydrodynamicsArbitrary2017}},
\begin{equation}
  \label{eq:switching_model}
\ten{\sigma}_{\text{swi}} = \nu \big( 1 - s(\alpha |\vec{B}|) \big) \ten{W} + \nu s(\alpha |\vec{B}|) \left[\frac{3}{2}(\ten{W}\vec{b}\cdot\vec{b}) \left( \vec{b} \otimes \vec{b} - \frac{1}{3}\ten{I} \right)\right],
\end{equation}
where $\vec{b} = \vec{B}/|\vec{B}|$ is the unit vector in the direction of the magnetic field and $s(\alpha |\vec{B}|)$ is the switching function, an interpolation function which controls the degree of anisotropy in the tensor based on the local magnetic field strength. This model focuses on the most important components (for the solar corona) of the full Braginskii tensor~\citep{braginskiiTransportProcessesPlasma1965}, the parallel and isotropic components, in order to better model viscosity in the vicinity of magnetic null points \citep{hollwegViscosityChewGoldbergerLowEquations1986a}. While previous work~\citep{mactaggartBraginskiiMagnetohydrodynamicsArbitrary2017} has used a phenomenological form for the \rs{switching}{} function $s$, here we use a form based on the coefficients of the Braginskii tensor,
\begin{equation}
  \label{eq:alt_switching1}
s(\rs{a}{x}) = \frac{3+f(2\rs{a}{x})-4f(\rs{a}{x})}{3},
\end{equation}
where
\begin{equation}
  \label{eq:eta_function}
  f(\rs{a}{x}) = \frac{6}{5}\frac{\rs{a}{x}^2 + 2.23}{\rs{a}{x}^4 + 4.03\rs{a}{x}^2 + 2.23},
\end{equation}
and $\rs{a}{x} = \alpha |\vec{B}|$, where $\alpha$ is a parameter used to control the dependence of $s$ on $|\vec{B}|$.
%  $x$ replaced by $a$ as $x$ is Cartesian coordinate

Physically, $\alpha = e\tau/m$ where $e$ is the electron charge, $m$ is the ion mass and $\tau$ is the ion-ion collision time
\begin{equation}
  \label{eq:collision_time}
  \tau = 0.82 \times 10^{-6} \frac{T^{3/2}}{n} \text{ s},
\end{equation}
where $n$ is the ion number density. For typical active region conditions, $T=2\times 10^6$ K and $n = 3 \times 10^3\text{ m}^{-3}$ giving $\tau = 0.773$ s and $\alpha \approx 10^{8}$. Were this value to be used in~\eqref{eq:alt_switching1}, $s$ would change so rapidly with $|\vec{B}|$ that the region near the null point where $s \approx 0$, corresponding to where the viscosity is isotropic, would be of sub-grid scale at the resolutions used here. To properly resolve the region of isotropic viscosity, we choose $\alpha = 12$. 

\subsection{Null point model}

The magnetic structure of the null point with the spine \rs{parallel to}{aligned along} the $z$-axis is \rs{determined by an imposed initial magnetic field given}{written} in non-dimensional units \rs{by}{as}
\begin{equation}
  \label{eq:null_point_field}
  \vec{B} = (x, y, -2z).
\end{equation}
The domain \rs{$\Omega$}{} is a box of dimension $[-3.5, 3.5]\times[-3.5, 3.5]\times [-0.25, 0.25] $ in the $x$, $y$ and $z$ directions, respectively. The initial velocity is uniformly zero, the initial density is uniformly $\rho = 1$ and the internal energy is uniformly $\varepsilon = 5/4$, corresponding to a temperature of $1.44 \times 10^9$ K and a plasma beta of $\beta \approx 0.017$.

\begin{figure}[t]
  \centering
  % RS - increased size to max
  \rs{\includegraphics[width=\linewidth]{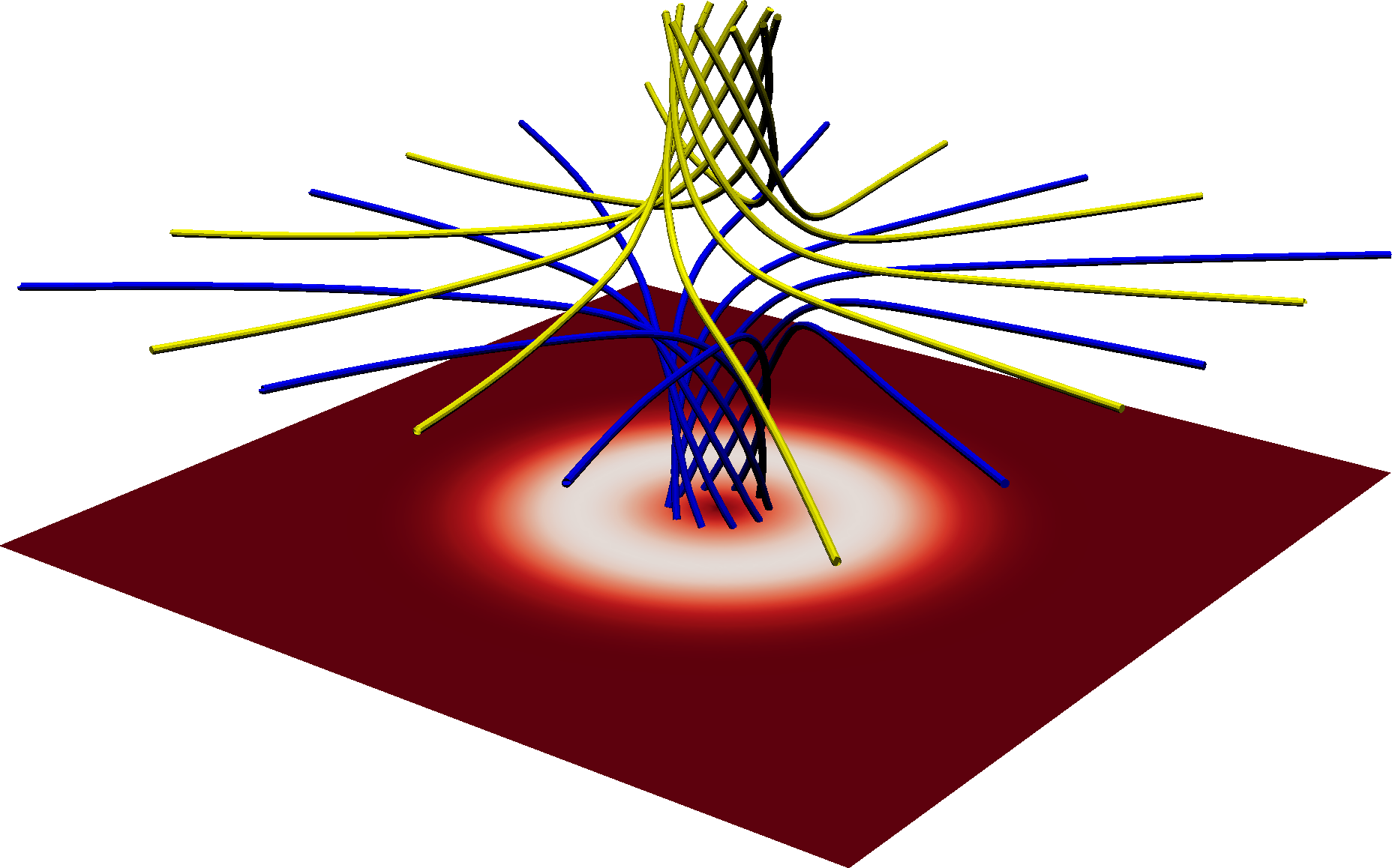}}
      {\includegraphics[width=0.7\linewidth]{field_line_plots/cropped/v-4r-4-isotropic_0008_cropped.png}}
  \mycaption{Field configuration after $4$ Alfv\'en times.}{The driver speed is also shown as a slice, where white indicates peak driving speed.}%
  \label{fig:field_line_plots/v-4r-4-iso-field-8}
\end{figure}

The driver takes the form of a slowly accelerating, rotating flow at the upper \rs{face of the box $z=0.25$ given by}{$z$-boundary as} 
\begin{equation}
  \label{eq:null_twisting_profile}
  \vec{u} = u_0\, u_r(r)\, u_t(t)\, \big(-y, x,0.25\big)^T.
\end{equation}
\rs{Here}{where} $u_r(r)$ describes the radial profile of the twisting motion in terms of the radius $r^2 = x^2 + y^2$,
\begin{equation}
  \label{eq:radial_twisting_function}
  u_r(r) = u_{r0}\,\big(1 + \tanh(1 - r_d r^2)\big),
\end{equation}
where $r_d$ controls the radial extent of the driver and $u_{r0}$ is a normalising factor. The extent of the driving region can be seen in figure~\ref{fig:field_line_plots/v-4r-4-iso-field-8}. \rs{The}{$u_t(t)$ describes the} imposed acceleration of the twisting motion \rs{is described by $u_t(t)$ with}{}
\begin{equation}
  \label{eq:ramping_up_function}
  u_t(t) = \tanh^2(t/t_r),
\end{equation}
where the parameter $t_r$ controls the time taken to reach the final driver velocity $u_0$. The parameters used in all simulations are $u_0 = 0.09$, $u_{r0} = 5.56$, $t_r = 0.25$ and $r_d = 36$. At the lower boundary \rs{$z=-0.25$}{} the flow is in the opposite direction.

This driver twists the footpoints of the spine of the null point, dragging the field and introducing twist throughout the entire null point (figure~\ref{fig:field_line_plots/v-4r-4-iso-field-8}). The form of driver allows the system to be accelerated slowly enough that the production of disruptive shocks and fast waves is minimal. It is unavoidable that some waves are produced during the boundary acceleration, however these provide a useful source of noise which acts as a perturbation. As in~\rs{\cite{wyperKelvinHelmholtzInstabilityCurrentvortex2013}}{\citet{wyperKelvinHelmholtzInstabilityCurrentvortex2013}}, there is no prescribed perturbation which results in either the KHI or the null collapse; all perturbations are dynamically generated due to noise in the system. This entire setup is similar to that of~\citet{wyperKelvinHelmholtzInstabilityCurrentvortex2013}. 

The main parameter study required 18 simulations to be run in total; one per viscosity model for each of the 9 parameter choices. To limit the time required to complete the study, a relatively low resolution of $320$ grid points in each direction is used for these runs. A single, higher-resolution pair of simulations were run, one for each viscosity model, at the resolution of $640$ grid points for a single parameter choice. As well as allowing a detailed analysis of this case, these higher-resolution simulations provide evidence that the lower resolution simulations have suitably converged.

\section{Methods of analysis}

\label{sec:khi_analysis}

\subsection{Stability measures}

\label{sec:stability_measures}

Following~\citet{wyperKelvinHelmholtzInstabilityCurrentvortex2013}, two quantities are used in understanding the stability of the current-vortex sheet: the fast mode Mach number $M_f$, associated with the velocity shear, and a parameter $\Lambda$ describing the balance of stability between the tearing mode and the KHI in a current-vortex sheet\footnote{\rs{\citet{wyperKelvinHelmholtzInstabilityCurrentvortex2013}}{Wyper and Pontin} additionally use the projected Alfv\'en Mach number alongside the two measures used here, however it is our opinion that $\Lambda$ captures the same information.}. The fast mode Mach number is given by
\begin{equation}
  \label{eq:mach_numbers}
  M_f = \frac{\Delta u}{\sqrt{c_s^2 + c_A^2}}\rs{,}{}
\end{equation}
where \rs{$\Delta u$ is the velocity change across the shear layer and}{} $c_s$ and $v_A$ are the local sound and Alfv\'en speeds, respectively. The parameter $\Lambda$ measures the relative strength of the velocity shear to magnetic shear and is given by
\begin{equation}
  \label{eq:khi_stability_param}
  \Lambda = \frac{L_b}{L_u} M_A^{2/3},
\end{equation}
where $M_A$ is the projected Alfv\'en Mach number
\begin{equation}
  \label{eq:alfven_mach_number}
M_A = \frac{\Delta u \sqrt{\rho}}{\Delta B}.
\end{equation}
Since the shear layer occurs in the presence of a guide field (that of the initial magnetic null point) which is not included in the linear stability study of the KHI, the difference in magnetic field $\Delta B$ \rs{across the shear layer and}{} is used in the Alfv\'en Mach number as opposed to the full magnetic field strength $|\vec{B}|$. In this way the Alfv\'en Mach number can be considered projected on to the shear layer. 

Plotting the radial dependence of these quantities over the shear layers gives an indication of the local linear stability based on the stability analysis performed by~\citet{einaudiResistiveInstabilitiesFlowing1986}. The analysis predicts that a current-vortex sheet is linearly unstable to the KHI where $M_f < 2$ and $\Lambda > 1$. Where $\Lambda < 1$, the analysis predicts that the sheet is unstable to the tearing instability instead. It should be noted that the analysis of~\citet{einaudiResistiveInstabilitiesFlowing1986} is 2D so can only be approximately used in the study of the KHI here, where there is an additional guide field in the system.

To calculate the stability measures, the peak vorticity and current density within the current-vortex sheets are measured, along with the radii at which the peaks occur. These radii are then used as the locations at which the absolute difference in azimuthal velocity $\Delta u$ and magnetic field $\Delta B$ across the shear layers are measured, calculated as the difference between the maximum and minimum values of velocity or magnetic field either side of the shear layer. The distance between the maximum and minimum points gives a measurement of the thickness of the shear layers, $L_u$ and $L_B$.

\subsection{Reconnection rate}
\label{sec:reconn_rate}

The reconnection rate is calculated using the same method employed in previous work by the same authors~\citep{quinnEffectAnisotropicViscosity2020a}. In summary, the reconnection rate local to a given magnetic field line is calculated as the local parallel electric field (that is, parallel to the magnetic field) integrated along the field line. By choosing a grid of starting points and integrating along each associated field line, an image is constructed of reconnection rates projected onto the grid of field line seed points. This is used to explore the spatial distribution of reconnection. The maximum value across all seed points gives the conventionally accepted measure of reconnection rate, the maximum integrated value~\citep{galsgaardSteadyStateReconnection2011,priestNatureThreedimensionalMagnetic2003,schindlerGeneralMagneticReconnection1988}.

\section{Results}

\label{sec:khi_results}

In the first \rs{subsection}{part} of the results \rs{section}{}, the evolution of \rs{a pair of high-resolution}{the high-resolution pair of} simulations is presented, both performed using resistivity of $\eta = 10^{-4}$ and viscosity $\nu = 10^{-4}$, 
\rs{but with different viscosity models}{and the effect of the two viscosity models are compared}. \rs{The purpose of these}{These} simulations \rs{is to}{} capture the main features of the \dm{null point dynamics}{ in the null point} in response to the driver: the formation of a current-vortex sheet in the fan plane, the appearance of counterflows, the (potential) growth of a KHI, and the eventual collapse of the null \rs{point}{}. This \rs{pair of high-resolution simulations}{simulation pair} also highlights\rs{, in detail, }{} the differences between the isotropic and the switching viscosity models, \rs{in particular}{mainly} the suppression of the KHI in the isotropic case, and the quicker collapse of the null \rs{point}{} in the switching case. These results are then \rs{extended}{generalised} to other parameter choices in \rs{the second subsection}{a later section}.

\subsection{Evolution of a typical case}
\label{sec:null_point_khi_single_case}

The evolution of the high-resolution, typical cases is detailed in stages, exploring the formation, stability and breakup of the current-vortex sheet before investigating the collapse of the null \rs{point}{}. Then, the evolution is summarised through an analysis of the energy budget and the reconnection rate in time.

\subsubsection{Formation of the current-vortex sheet}

\begin{figure}[t]
  \centering
\begin{subfigure}{\linewidth}
      \includegraphics[width=\linewidth]{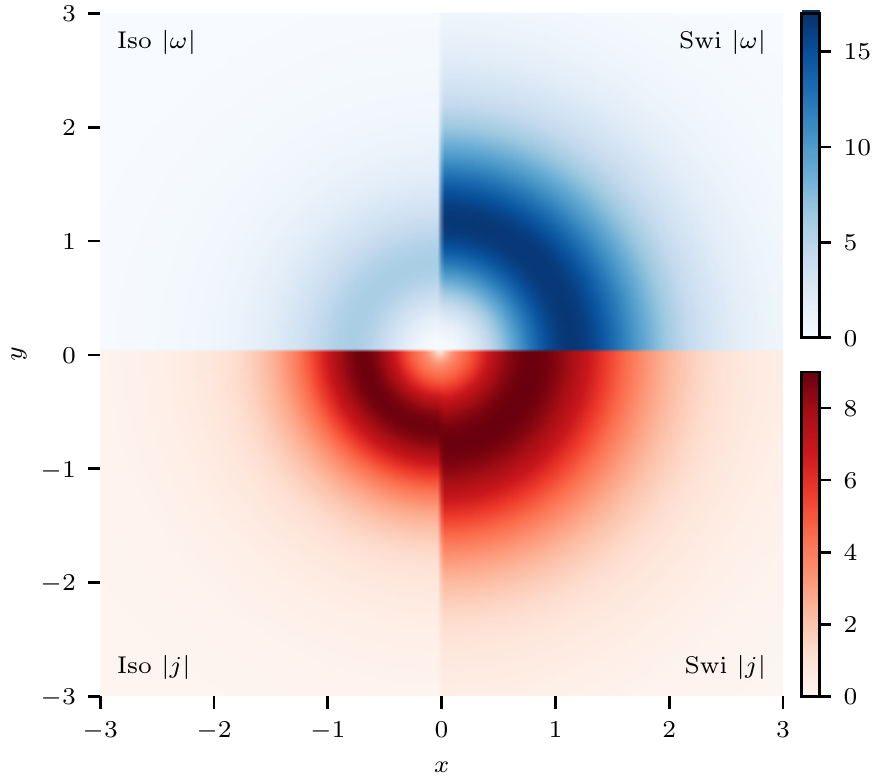}
      \caption{Vorticity and current density rings.}
      \label{fig:v-4r-4_vorticity_current_ring_t_3}
      \vspace{4mm}
    \end{subfigure}
    \hfill
\begin{subfigure}{0.96\linewidth}
    \hspace{-3mm}
      \includegraphics[width=\linewidth]{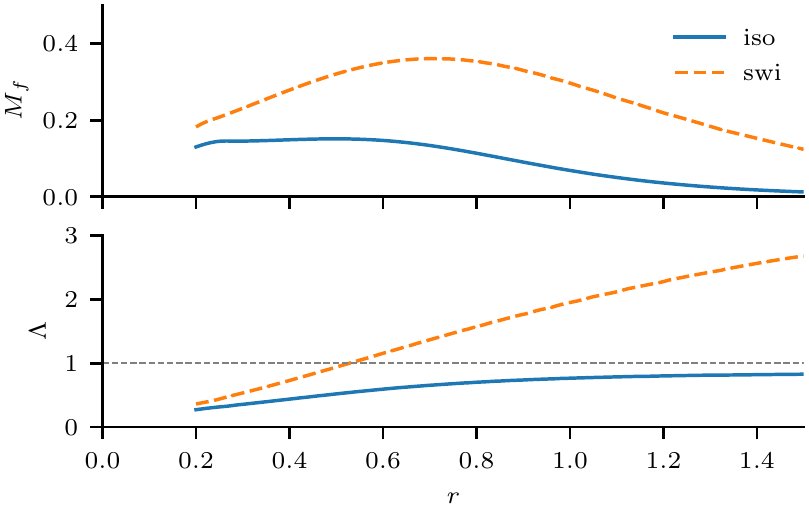}
      \caption{Stability measures}%
      \label{fig:v-4r-4_mach_t_6}
    \end{subfigure}
\mycaption{Rings of vorticity and current density and associated linear stability criteria.}{Subfigure (a) plots the vorticity and current density for both viscosity models at $t=3$ \rs{and $z=0$}{}. Subfigure (b) plots the linear stability measures as functions of radius at $t=6$. The switching model permits rings of greater radial extent and notably stronger vorticity resulting in a current-vortex sheet which is linearly unstable to the KHI. \rs{The thin dashed grey line $\Lambda=1$ is the boundary between the KHI and the tearing instability from the 2D linear analysis of \citet{einaudiResistiveInstabilitiesFlowing1986}}{}.}
\label{fig:rings_and_stability}%
\end{figure}

Initially, the torsional Alfv\'en waves injected by the driver trace \rs{out}{} the field surrounding the null, moving first along the spine then out across the fan plane. This occurs from above and below. The upper and lower waves diffuse via viscosity and resistivity and eventually meet, creating shear layers in the velocity and magnetic field in the form of rings of vorticity and current density centred around the null point (figure~\ref{fig:v-4r-4_vorticity_current_ring_t_3}). These shear layers are jointly called the current-vortex sheet. Without any diffusion in the system the waves would travel far along the fan plane before meeting. The presence of both viscosity and resistivity diffuses the waves as they travel along the field, allowing the upper and lower waves to meet around $r=1$, where the current-vortex sheet forms. The hole in the sheet is due to magnetic tension forces opposing the twisting motion, as illustrated in figure 3 of~\rs{\cite{wyperKelvinHelmholtzInstabilityCurrentvortex2013}}{\citet{wyperKelvinHelmholtzInstabilityCurrentvortex2013}}. This also gives rise to counterflows (not shown) similar to those seen in~\rs{\cite{wyperKelvinHelmholtzInstabilityCurrentvortex2013}}{\citet{wyperKelvinHelmholtzInstabilityCurrentvortex2013,galsgaardNumericalExperimentsWave2003}} \dm{and \cite{galsgaardNumericalExperimentsWave2003}}{}.

In the switching case, the reduced effective viscosity produces a vortex ring which is larger in radius and stronger in magnitude. The current density ring is somewhat larger in the switching case, but of equivalent peak magnitude to that in the isotropic case. Since the viscosity diffuses velocity directly and affects the magnetic field only indirectly, the vorticity is naturally affected by the change in viscosity model more than the current density. This difference in vorticity but not current density affects the relative size of the stability measures.

\begin{figure*}[t]
  \centering
% rs - max out fig size   
%\begin{subfigure}{0.7\linewidth}  
\begin{subfigure}{0.32\linewidth}
      \includegraphics[width=\linewidth]{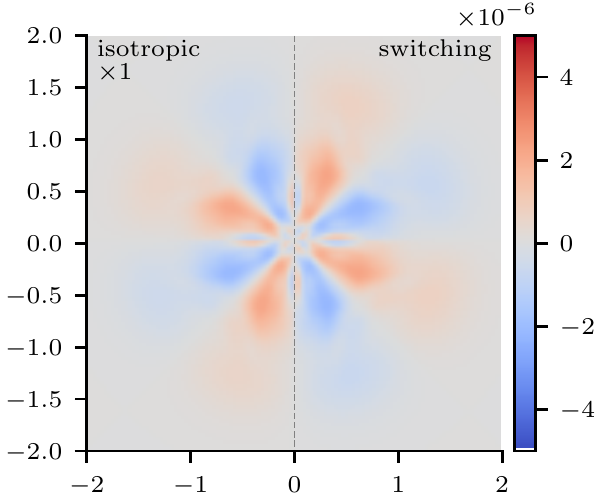}
      \caption{$t=2$}
      \label{fig:v-4r-4_uz_t_2}
    \end{subfigure}
% rs - max out fig size   
%\begin{subfigure}{0.7\linewidth}     
\begin{subfigure}{0.34\linewidth}
\hspace{1mm}
      \includegraphics[width=\linewidth]{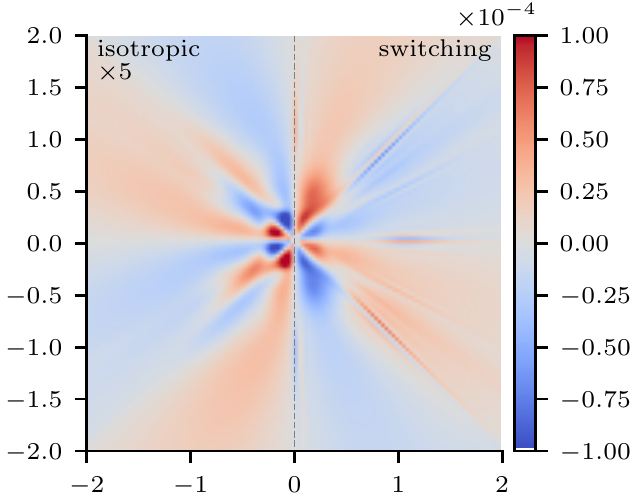}
      \caption{$t=6$}
      \label{fig:v-4r-4_uz_t_6}
    \end{subfigure}
% rs - max out fig size   
%\begin{subfigure}{0.7\linewidth} 
\begin{subfigure}{0.32\linewidth}
      \includegraphics[width=\linewidth]{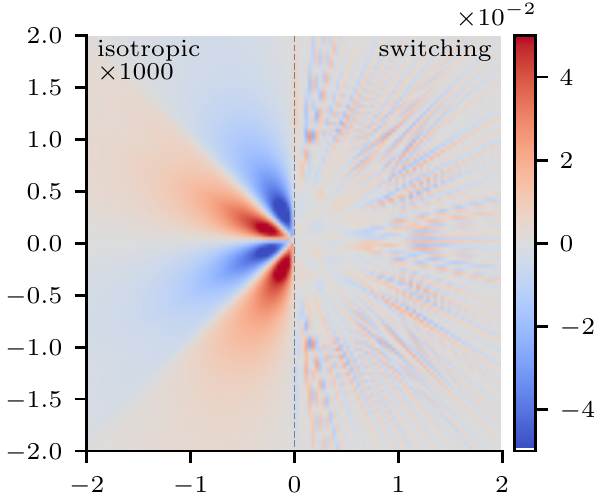}
      \caption{$t=10$}
      \label{fig:v-4r-4_uz_t_10}
    \end{subfigure}
\mycaption{Development of the KHI in the \rs{out-of-plane}{out of plane} velocity $u_z$ at $t=2, 6$ and $10$ for both viscosity models.}{Note the isotropic results have been multiplied by as much as $1000$ in order to compare to the switching results. In the switching case the KHI appears initially along the diagonals before extending azimuthally. In the isotropic case there is no evidence of the instability.}
\vspace{-2mm}
\label{fig:out_of_plane_velocity}%
\end{figure*}

Figure~\ref{fig:v-4r-4_mach_t_6} shows the relevant stability measures as functions of radius across the fan plane at $t=6$, a time when the fan plane has become unstable to the KHI in the switching case but remains stable in the isotropic case (see figure~\ref{fig:v-4r-4_uz_t_6}). The measure $\Lambda$ confirms that the current-vortex sheet is linearly stable to the KHI in the isotropic case and unstable in the switching case for $r>0.6$. This linear prediction matches \rs{the location}{} where the KHI is observed to develop. In the switching case the peak of $M_f$ aligns with the observed region of initial growth of the instability.

In the switching case, $\Lambda$ and $M_f$ are significantly larger due to the greater vorticity (figure~\ref{fig:v-4r-4_vorticity_current_ring_t_3}). In the isotropic case the more efficient dissipation of velocity results in a generally weaker vorticity ring and, thus, lower \rs{values of the}{} stability measures.

\subsubsection{Disruption of the current-vortex sheet}

As expected from the linear stability measures shown in figure~\ref{fig:v-4r-4_mach_t_6}, figure~\ref{fig:out_of_plane_velocity} shows the development of the \rs{out-of-plane}{out of plane} velocity from $t=2$ to $10$ and reveals that the current-vortex sheet in only the switching case is unstable to the KHI. Both cases exhibit a similar morphology of $u_z$, despite a growing difference in strength, until $t=6$ when the KHI appears only in the switching case, initially along the diagonals (figure~\ref{fig:v-4r-4_uz_t_6}) before spreading azimuthally (figure~\ref{fig:v-4r-4_uz_t_10}). There is no evidence of the KHI in the isotropic case.

\begin{figure}[t]
  \centering
% \begin{subfigure}{0.75\linewidth} % For referee version
\begin{subfigure}{\linewidth}
      \includegraphics[width=\linewidth]{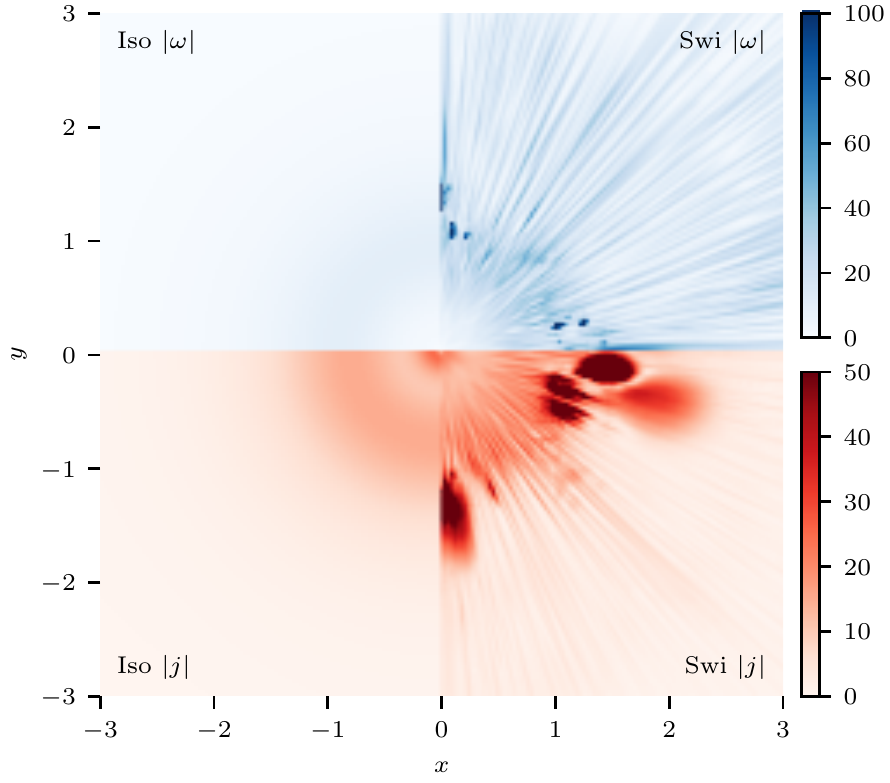}
      \caption{Current-vortex sheet}
      \label{fig:v-4r-4_vorticity_current_ring_t_10}
    \end{subfigure}
    \vspace{4mm}
    \hfill
    % \begin{subfigure}{0.75\linewidth} % For referee version
    \begin{subfigure}{\linewidth}
  \includegraphics[width=\linewidth]{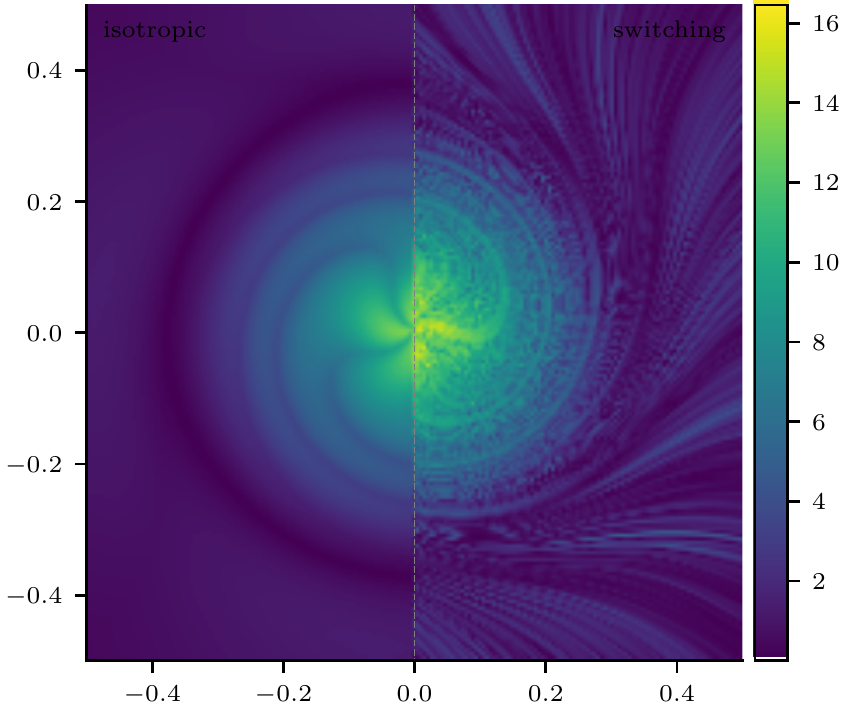}
      \caption{Reconnection rate}
      \label{fig:v-4r-4_reconn_rate_t_10}
    \end{subfigure}
\mycaption{The breakup of the current-vortex sheet and associated reconnection at $t=10$ \rs{and $z=0$}{}.}{Subfigure (a) presents the current and vorticity density and subfigure (b) presents the spatial distribution of reconnection rate measured as the parallel electric field integrated along field lines traced from locations in the plane $z=0.23$. Both subfigures show both viscosity models with isotropic shown on the left half of each figure and switching on the right. The current-vortex sheet remains stable in the isotropic case while that in the switching case has been fragmented by the KHI. The resultant small-scale reconnection in the rolls produce localised pockets of strong vorticity and current density.}
\end{figure}

In both cases the current-vortex sheet grows in radius and magnitude with time, more in the switching case than in the isotropic. The shearing action of the counterflows produces a secondary ring of strong current density closer to the spine which is greater in magnitude in the isotropic case. By $t=10$ the KHI has disrupted the current-vortex sheet (figure~\ref{fig:v-4r-4_vorticity_current_ring_t_10}) and the resultant rolls create strong, small-scale current sheets, enhancing the local reconnection rate. 

Figure~\ref{fig:v-4r-4_reconn_rate_t_10} shows the spatial distribution of the reconnection rate for both viscosity models. Each pixel in the image represents one field line passing through that pixel along which the parallel electric field has been integrated. The colour of the pixel is given by the value of the integration. The reconnection rate is greatest close to the origin, corresponding to regions of slippage reconnection due to the strong currents in the spine and current-vortex sheet. The effects of the boundary can be seen as long dark lines which spiral outwards from the origin. The switching case shows a greater peak reconnection rate due to the small scale current sheets created by the KHI, and the enhanced reconnection far from the null can be seen as ripple-like structures in the fringes of the plot.

\subsubsection{Spine-fan reconnection}

This section presents the results of driving the magnetic null \rs{point}{} to the \rs{moment}{point} at which it undergoes spontaneous collapse. The collapse is instigated by a velocity shear across the null which generates a magnetic shear, permitting spine-fan reconnection. The results of the isotropic case are presented first, in detail, then the effect of the KHI is explored in the switching case.

\begin{figure}[t]
  \centering
% rs - max out fig size
%    \begin{subfigure}{0.70\linewidth}
\begin{subfigure}{\linewidth}
      \includegraphics[width=\linewidth]{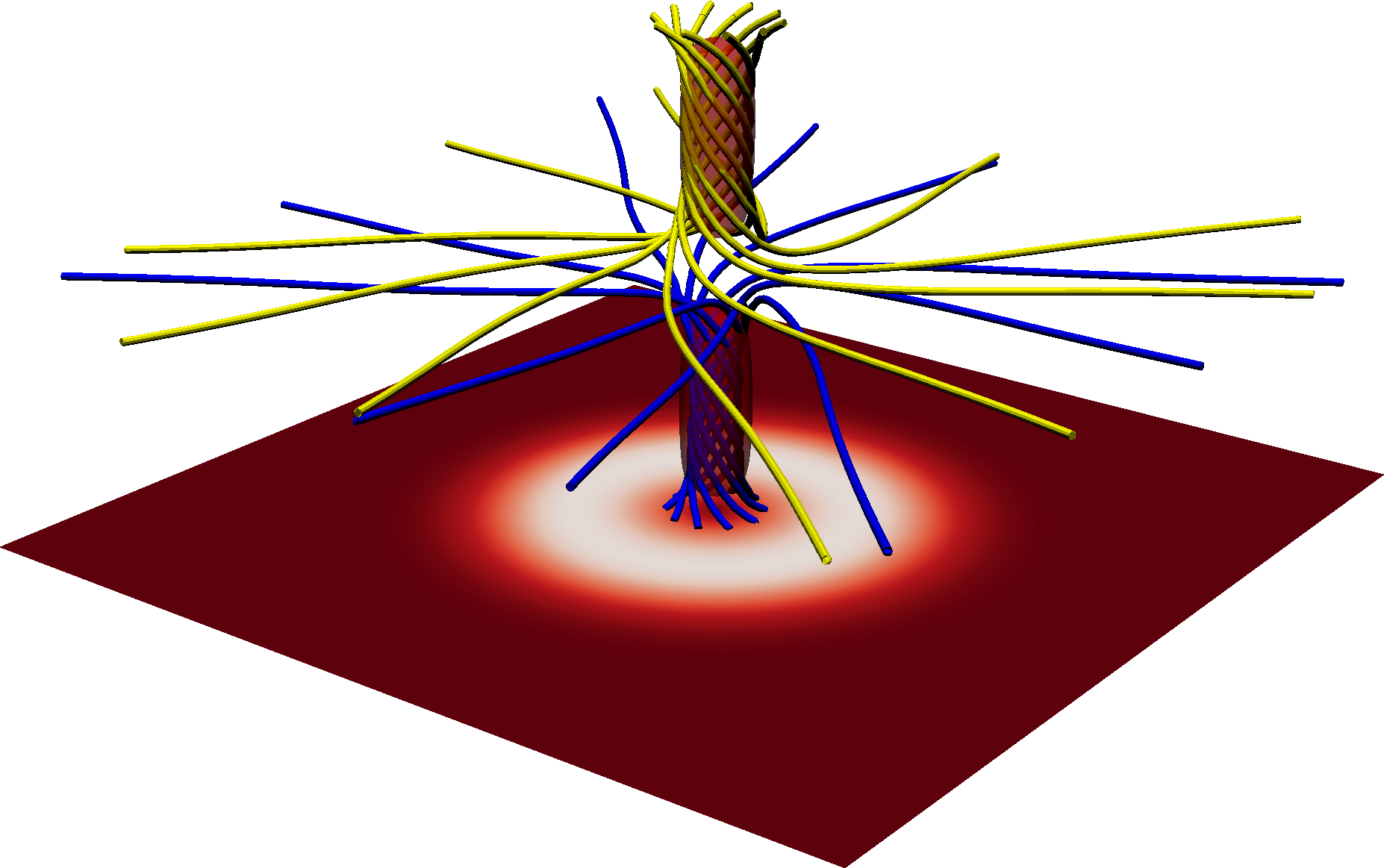}
      \caption{$t=15$}
      \label{fig:v-4r-4-iso-field-30}
    \end{subfigure}
    \hfill
    \vspace{4mm}
% rs - max out fig size
%\begin{subfigure}{0.70\linewidth}
\begin{subfigure}{\linewidth}
      \includegraphics[width=\linewidth]{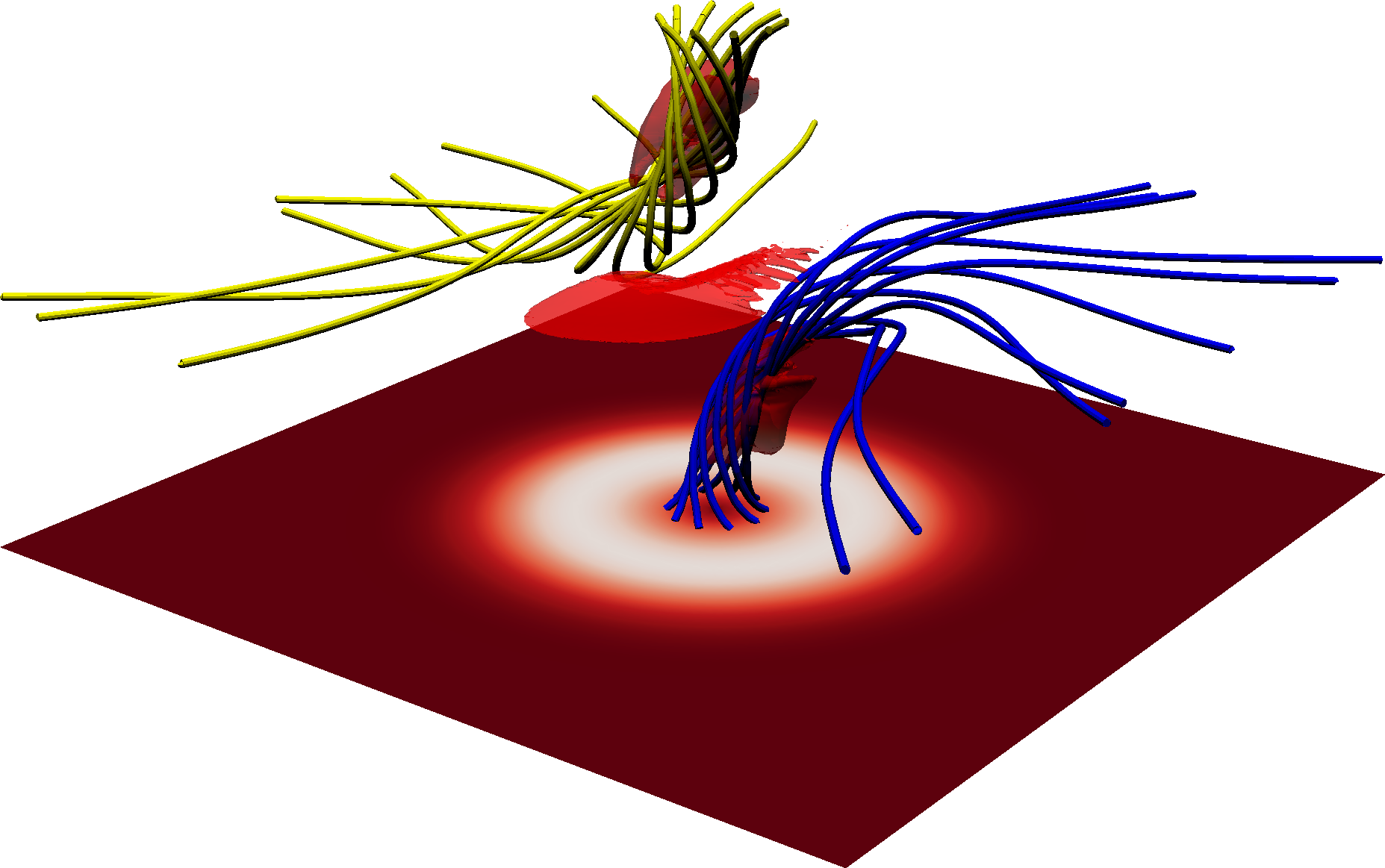}
      \caption{$t=18.5$}
      \label{fig:v-4r-4-iso-field-37}
    \end{subfigure}
\mycaption{Collapse of the null point visualised with field lines in the isotropic case.}{Field lines are plotted from a circle of radius $0.05$ around the upper and lower spine footpoints. Contours of $|\vec{j}| = 60$ are also plotted and reveal the strong current within the spine as well as the formation of the central sheet associated with the spine-fan reconnection. At $t=18.5$ the bulk of the field lines making up the core of the spine have reconnected.}
\label{fig:khi_field_lines_collapse}
\end{figure}

In typical studies of spine-fan reconnection (such as~\citet{pontinCurrentSheetFormation2007}) the spines of a null point are dragged in opposite directions at the boundaries. This motion pulls the field above and below the null point in opposite directions and creates a current sheet which acts to reconnect field lines between the spine and fan. Here, the field near the null is shifted not because of motions at the footpoints of the spine, but due to imbalances in the velocity above and below the null point which arise due to small pressure differences generated during the course of the initial driving. Figure~\ref{fig:khi_field_lines_collapse} presents the magnetic field lines before and during the reconnection.

\begin{figure}[t]
  \centering
    \begin{subfigure}{0.56\linewidth}
      \includegraphics[width=\linewidth]{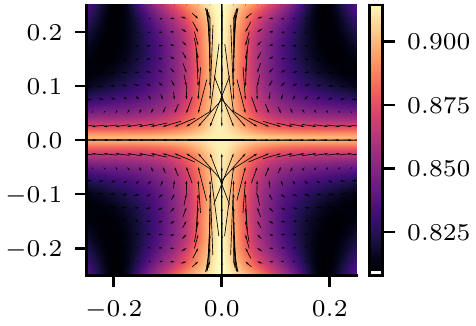}
      \caption{Pressure and flow}
      \label{fig:v-4r-4-pressure-flow-30}
    \end{subfigure}
    \hfill
    \begin{subfigure}{0.43\linewidth}
      \includegraphics[width=\linewidth]{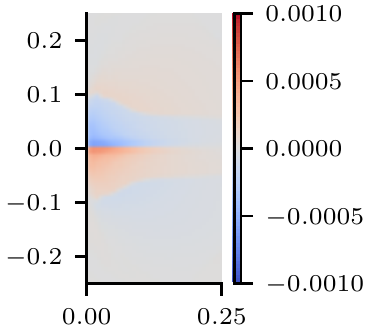}
      \caption{Imbalance in $u_x$}
      \label{fig:v-4r-4-vx-imbalance-30}
    \end{subfigure}

\mycaption{Velocity imbalance above and below the null \rs{point}{}.}{Subfigure (a) plots a slice of the pressure through $y=0$ overlaid with fluid velocity, where the longest arrows correspond to a fluid velocity of approximately $0.1$. Subfigure (b) depicts $|u_x(x)| - |u_x(-x)|$, the difference in $u_x$ between the left and right sides of the plane $x=0$. This gives a measure of the asymmetry in the velocity around the null point.}
\label{fig:imbalance_in_velocity}
\end{figure}

The twist in the spines creates a current which heats the contained plasma via \rs{ohmic}{Ohmic} heating and generates a small pressure force directed towards the null point, driving two oppositely directed streams of plasma along the spine (figure~\ref{fig:v-4r-4-pressure-flow-30}). Where these streams meet (at the null point) they form a stagnation point flow, compressing the plasma in the vicinity of the null \rs{point}{} and flowing out along the fan plane. Due to small asymmetries in the pressure that accrue during the simulation, an imbalance in the velocity appears above and below the null point (figure~\ref{fig:v-4r-4-vx-imbalance-30}). 

The velocity shear around the null \rs{point}{} shears the magnetic field accordingly, creating a current sheet through the null point (figure~\ref{fig:v-4r-4-iso-field-37}). This current sheet enables reconnection between the spine and fan which further extends, thins and strengthens the sheet, continuing the reconnection process until the field around the null \rs{point}{} collapses. The collapse itself can be seen in the kinetic energy as a dramatic increase starting at $t\approx18$ (figure~\ref{fig:v-4r-4_kinetic_energy}). The development of the current sheet and the resultant spine-fan reconnection is similar to that of~\citet{pontinCurrentSheetFormation2007} with the exception that the twist in the field unravels as the reconnection proceeds. In the switching case, the development of the spine-fan reconnection and associated collapse is qualitatively similar to that in the isotropic case with the exception that the reconnection occurs notably earlier and evolves over a shorter timescale.

\subsubsection{\rs{Energy}{Development of energy} budget and reconnection rate}

\begin{figure*}[t]
  \centering
  \begin{subfigure}{0.33\linewidth}
    \includegraphics[width=\linewidth]{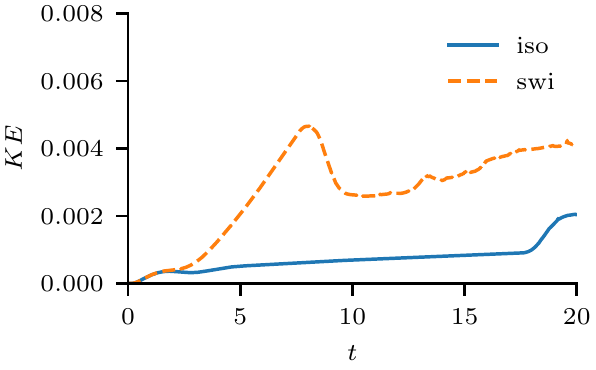}
    \caption{Kinetic energy}
    \label{fig:v-4r-4_kinetic_energy}
  \end{subfigure}
    \hfill
  \begin{subfigure}{0.33\linewidth}
   \hspace{-2mm}\raisebox{3mm}{ \includegraphics[width=\linewidth]{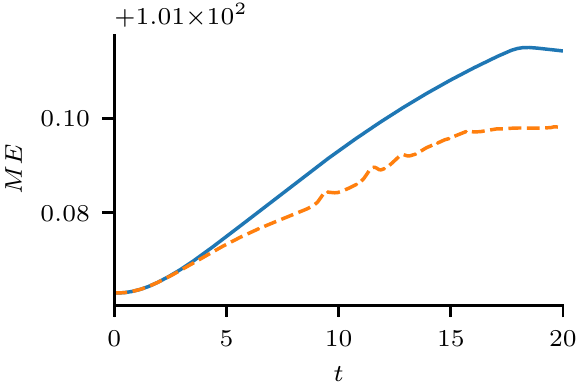}}
    \caption{\rs{Magnetic energy}{}}%
    \label{fig:v-4r-4_magnetic_energy}
  \end{subfigure}
  \begin{subfigure}{0.33\linewidth}
  \hspace{-3mm}  \includegraphics[width=\linewidth]{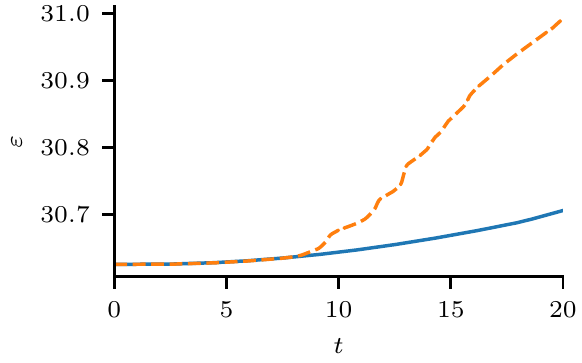}
    \caption{Internal energy}%
    \label{fig:v-4r-4_internal_energy}
  \end{subfigure}
  \hfill
  \begin{subfigure}{0.33\linewidth}
    \includegraphics[width=\linewidth]{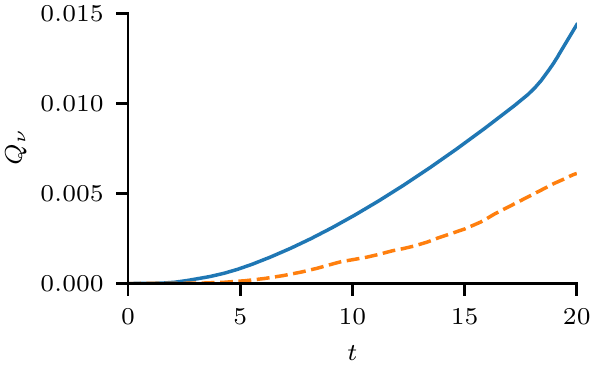}
    \caption{Viscous heating}%
    \label{fig:v-4r-4_viscous_heating}
  \end{subfigure}
  \begin{subfigure}{0.33\linewidth}
    \includegraphics[width=\linewidth]{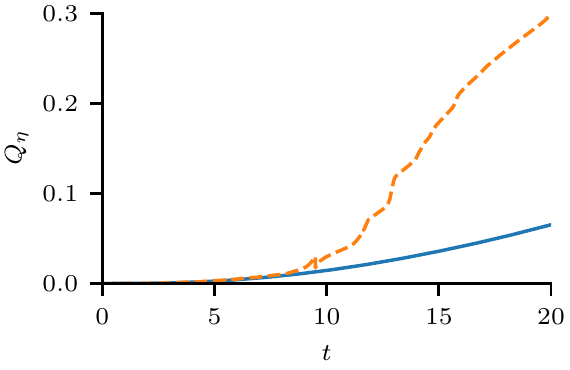}
    \caption{\rs{Ohmic heating}}%
    \label{fig:v-4r-4_ohmic_heating}
  \end{subfigure}
  \hfill
  \begin{subfigure}{0.33\linewidth}
    \includegraphics[width=\linewidth]{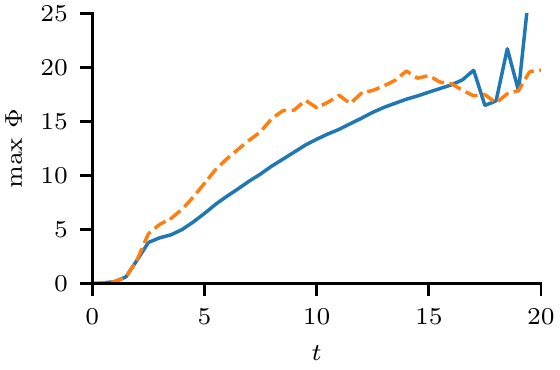}
    \caption{Reconnection rate}%
    \label{fig:v-4r-4_reconn_rate_over_time}
  \end{subfigure}
  \mycaption{Energy \rs{components}{measures} and reconnection rate as functions of time.}{}
\end{figure*}

The kinetic energy in the switching case \rs{is a measure of}{describes} the main evolution of a KHI-unstable current-vortex sheet as presented in detail previously (figure~\ref{fig:v-4r-4_kinetic_energy}). The initial injection of the Alfv\'en waves and formation of the current-vortex sheet can be seen at at $t\approx3$. As the null \rs{point}{} continues to be driven, the sheet becomes unstable to the KHI and the kinetic energy grows accordingly from $t\approx3$ to $8$. At $t\approx 8$ the KHI saturates as small-scale current sheets form and Ohmic heating begins to drain energy from the instability (figure~\ref{fig:v-4r-4_ohmic_heating}). This is also reflected in the reconnection rate (figure~\ref{fig:v-4r-4_reconn_rate_over_time}) where the small current sheets in the rolls of the KHI in the fan plane enhance reconnection locally. Around $t=14$, a transient increase in the kinetic energy reveals the start of the null \rs{point}{} collapse. 

In the isotropic case, the increased kinetic energy and enhanced reconnection rate associated with the KHI are absent, however the collapse of the null \rs{point}{} produces significantly more kinetic energy at $t\approx 17$ than in the switching case (figure~\ref{fig:v-4r-4_kinetic_energy}). The \rs{ohmic}{Ohmic} heating is similarly damped without the influence of the KHI (figure~\ref{fig:v-4r-4_ohmic_heating}). This results in the switching model extracting more energy from the field (figure~\ref{fig:v-4r-4_magnetic_energy}) and heating the plasma much more effectively (figure~\ref{fig:v-4r-4_internal_energy}). One significant finding is that the velocity shears created by the KHI allow anisotropic viscous heating of comparable levels to that of isotropic viscosity (in contrast to \rs{much larger differences}{the orders of magnitude difference} observed in other \rs{situations e.g.~the kink instability of a twisted flux tube \citep{quinnEffectAnisotropicViscosity2020a}}{chapters}). 

The reconnection rate reveals some interesting features about the nature of reconnection within the system and how the presence of the KHI affects the null \rs{point}{} collapse (figure~\ref{fig:v-4r-4_reconn_rate_over_time}). One interesting observation is that the switching case shows a greater reconnection rate than that of the isotropic case even before the onset of the KHI (i.e. for $t < 6$), suggesting the switching model \dm{allows for enhanced}{itself is enhancing} reconnection. As in \cite{quinnEffectAnisotropicViscosity2020a}, here too, this is due to the switching model permitting greater velocities, greater compression and thinner, stronger current sheets. It is then unclear whether the generally enhanced reconnection rate in the switching case for times $t=5$ to $10$ can be attributed to the current-enhancing effect of the switching model or an effect of the KHI. Certainly, the spiky nature of the reconnection rate from $t=8$ to $15$ can be attributed to the small, strong current sheets produced in the rolls of the KHI, which do not appear in the stable isotropic case. The collapse of the null \rs{point}{} is observed in the reconnection rate in the switching case around $t=15$ and in the isotropic case around $t=17$, however it differs significantly between the two cases. In the isotropic case, the reconnection rate increases during the collapse, while in the switching case, it decreases. This is due to the difference in the state of the nulls in each case as the collapse occurs.

In the isotropic case, where the KHI has not been excited, the flows and magnetic field are relatively simple and smooth such that the collapse is able to form large current sheets and reconnect many field lines at once. In contrast, in the switching case the KHI has broken up the current-vortex sheet, introduced inhomogeneities throughout the fan plane and generated small current sheets. This results in a collapse which struggles to reconnect with the same efficiency as in the smoother, simpler isotropic case. Additionally, there is simply more free magnetic energy in the system where the KHI remains stable, allowing current sheets to form more effectively during the collapse. In essence, the KHI places the null \rs{point}{} in a more complex state where the collapse is less efficient at reconnecting field lines.

\subsection{\rs{Study of parameter dependencies}{Analysis of parameter study}}

The results shown in section~\ref{sec:null_point_khi_single_case} change dramatically when \rs{the resistivity}{} $\nu$ and \rs{the viscosity}{} $\eta$ are varied. This section presents results of simulations where $\nu$ \rs{takes values of}{is varied as} $10^{-5}$, $10^{-4}$ or $10^{-3}$ and $\eta$ \rs{takes values of}{as} $10^{-4}$ or $10^{-3}$. This results in six pairs of simulations, each choice being run with switching viscosity or isotropic viscosity. The simulations were performed at a resolution of $320$ grid points per dimension, half that of the \rs{high-resolution cases described in detail above}{typical cases}, and run to $t=15$ instead of \rs{to}{} $t=20$ as in the \rs{latter}{typical cases}. The null \rs{point}{} collapse occurs sooner than at higher resolution and shows behaviour more typical of fast reconnection indicative of inadequate resolution~\citep{miyamaNumericalAstrophysicsProceedings2012}. For this reason, focus is placed on the development of the KHI rather than \rs{on}{} the null \rs{point}{} collapse, leaving a parameter study of the null \rs{point}{} collapse itself open as an avenue of future research. Generally, increasing \rs{the resistivity}{} $\eta$ to $10^{-3}$ produces a null \rs{point}{} that is more unstable to the KHI (even in isotropic cases). Increasing \rs{the viscosity}{} $\nu$ damps the KHI but does not totally suppress it, while decreasing $\nu$ leads to a more unstable KHI.

\subsubsection{Shear layer stability}

%\begin{figure}[t]
    %\begin{subfigure}{0.49\textwidth}
      %\centering
  %\includegraphics[width=1.0\linewidth]{param_study/peak_vort.pdf}
      %\caption{Peak vorticity}%
      %\label{fig:peak_vort}
    %\end{subfigure}
    %\hfill
    %\begin{subfigure}{0.49\textwidth}
      %\centering
  %\includegraphics[width=1.0\linewidth]{param_study/peak_current.pdf}
      %\caption{Peak current}%
      %\label{fig:peak_current}
    %\end{subfigure}
  %\mycaption{Peak vorticity and current as functions of viscosity $\nu$ for each value of resistivity $\eta$ at $t=8$.}{In the isotropic case, both rings decrease in radial extent as either diffusion parameter is increased. In the switching case, both rings also decrease with $\eta$, however there is a notable increase in the radial extent with $\nu$, particularly for high values of $\eta$.}%
  %\label{fig:param_study_peak_mag_and_loc}
%\end{figure}

\begin{figure*}[t]
\hfill
    \begin{subfigure}{0.26\linewidth}
      \centering
  \includegraphics[width=1.0\linewidth]{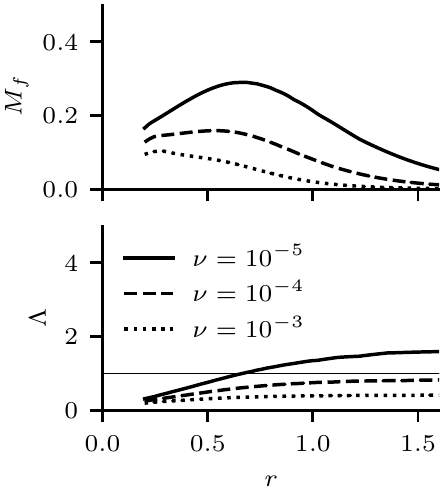}
      \caption{Isotropic; $\eta = 10^{-4}$}%
      \label{fig:mach_numbers_eta_4_iso}
    \end{subfigure}
    \hfill
    \begin{subfigure}{0.23\linewidth}
      \centering
  \includegraphics[width=1.0\linewidth]{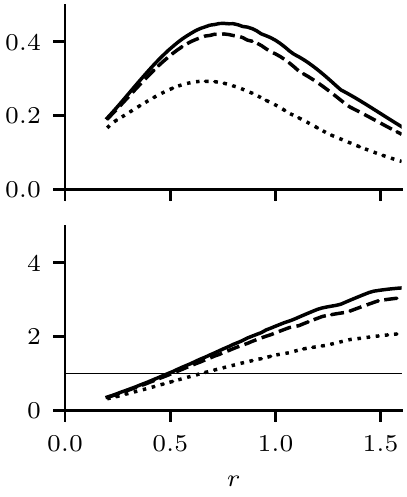}
      \caption{Switching; $\eta = 10^{-4}$}%
      \label{fig:mach_numbers_eta_4_swi}
    \end{subfigure}
    \hfill
    \begin{subfigure}{0.23\linewidth}
      \centering
  \includegraphics[width=1.0\linewidth]{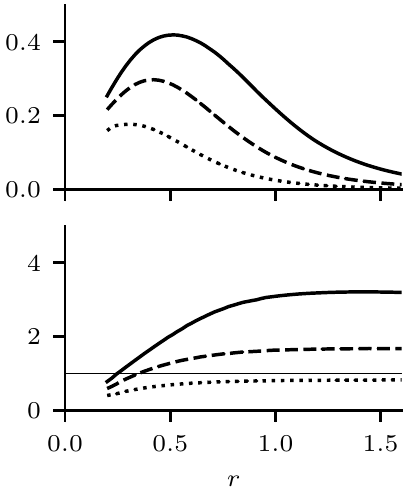}
      \caption{Isotropic; $\eta = 10^{-3}$}%
      \label{fig:mach_numbers_eta_3_iso}
    \end{subfigure}
    \hfill
    \begin{subfigure}{0.23\linewidth}
      \centering
  \includegraphics[width=1.0\linewidth]{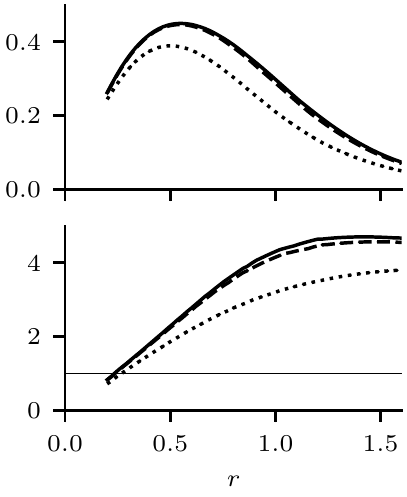}
      \caption{Switching; $\eta = 10^{-3}$}%
      \label{fig:mach_numbers_eta_3_swi}
    \end{subfigure}
  \mycaption{Plots of stability measures as functions of radius $r$ for all parameter choices at $t=8$.}{Measures are plotted for $\nu=10^{-5}$ (solid), $\nu=10^{-4}$ (dashed) and $\nu=10^{-3}$ (dotted) for each value of $\eta$ and each viscosity model. The cases where $\Lambda > 1$ are unstable with the exception of the isotropic cases where $\nu=10^{-4}$,  $\eta=10^{-3}$ (dashed line in figure~\ref{fig:mach_numbers_eta_3_iso}).}%
  \label{fig:mach_numbers}
\end{figure*}

Figure~\ref{fig:mach_numbers} \rs{shows}{plots} the stability measures as functions of radius for every studied parameter choice and \rs{for}{} both viscosity models at $t=8$. In every case $M_f < 2$, a necessary condition for \rs{instability of the}{an unstable}
current-vortex sheet. The condition on $\Lambda$ for instability is $\Lambda > 1$. All layers \dm{with the switching model}{} are linearly unstable to the KHI (figures~\ref{fig:mach_numbers_eta_3_swi} and~\ref{fig:mach_numbers_eta_4_swi}) while the isotropic cases show a mix of linear stability. When $\nu = 10^{-5}$, the viscosity is weaker and the linear stability analysis predicts that the layers should be unstable for either value of $\eta$. The opposite is true for $\nu=10^{-3}$, when the isotropic viscosity is at its most dissipative. The two middle cases, when $\nu=10^{-4}$ show stability when $\eta=10^{-4}$ and instability when $\eta=10^{-3}$. These findings can be attributed to the way in which diffusion affects the magnitude and thickness of the current-vortex sheet.

In general, increased diffusion leads to a thicker, weaker ring, due to the Alfv\'en waves diffusing more before meeting in the fan plane. The switching model, being generally less diffusive than the isotropic model, permits velocity shear layers with much greater peak vorticity. Due to the coupling between the magnetic field and the velocity in an Alfv\'en wave, the isotropic model appears to provide some diffusion to the magnetic field during the formation of the magnetic shear layer, resulting in a layer with weaker peak current, however the switching model affects the magnetic layer very little. Lower resistivity results in a stronger vorticity layer (in the switching case) but also a stronger current layer, and \emph{vice versa} for larger resistivity.

\begin{table}[t]
\centering
\begin{tabular}{llllllll}
$\eta$    & $\nu$     & Iso lin & Iso obs & Swi lin & Swi obs \\
\midrule
$10^{-3}$ & $10^{-3}$ & Stable     & Stable       & Unstable   & Unstable         & \\
$10^{-3}$ & $10^{-4}$ & Unstable   & Stable       & Unstable   & Unstable         & \\
$10^{-3}$ & $10^{-5}$ & Unstable   & Unstable     & Unstable   & Unstable         & \\
$10^{-4}$ & $10^{-3}$ & Stable     & Stable       & Unstable   & Unstable*         & \\
$10^{-4}$ & $10^{-4}$ & Stable     & Stable       & Unstable   & Unstable         & \\
$10^{-4}$ & $10^{-5}$ & Unstable   & Unstable*     & Unstable   & Unstable         &
\end{tabular}
\mycaption{\rs{Stability in the isotopic and switching cases for different choices of $\nu$ and $\eta$.}{}}{Both linear stability (\rs{"lin"}{}, as predicted by $\Lambda > 1$ in figure~\ref{fig:mach_numbers}) and observed stability \rs{("obs")}{} are shown. Entries marked as unstable* show growth of the KHI but the growth rate of the perturbation is close to zero. The isotropic model mostly results in stability while the switching model mostly results in instability.}
\label{tab:stability}
\end{table}

The observed stability of the current-vortex sheet to the KHI in each case is determined via inspection of the \rs{out-of-plane}{out of plane} velocity for each parameter choice and is summarised for each parameter choice in table~\ref{tab:stability}. Some entries are marked as unstable*, referring to their being marginal cases, that is the KHI is directly observed in the \rs{out-of-plane}{out of plane} velocity but the growth rate is close to zero and the perturbation remains negligibly small even at the final time of $t=15$. The observed stability is well matched by the theoretical conditions of instability $\Lambda > 1$ and $M_f < 2$ in all but one case. \rs{This exception at $\eta=10^{-3}$, $\nu=10^{-4}$ is close to marginal and the result is most likely affected by the choice of resolution}{}. This indicates that, despite the difference in geometry, the stability analysis of~\citet{einaudiResistiveInstabilitiesFlowing1986} is of practical use in predicting the stability of the KHI in magnetic null points. This condition even accurately predicts the stability of the marginal cases. 

\begin{figure*}[t]
  \centering

    \hfill
    \begin{subfigure}{0.33\linewidth}
      \centering
      \includegraphics[width=1.0\linewidth]{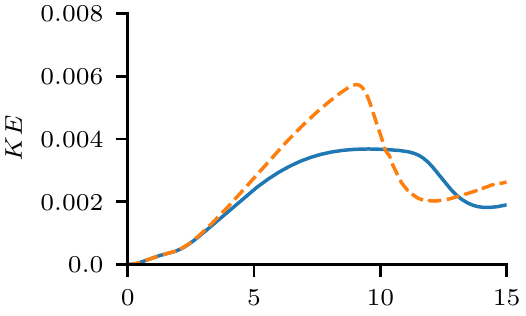}
      \caption{$\nu = 10^{-5};\ \eta = 10^{-3}$}%
      \label{fig:v-5r-3_kinetic_energy_ps}
    \end{subfigure}
    \hfill
    \begin{subfigure}{0.33\linewidth}
      \centering
      \includegraphics[width=1.0\linewidth]{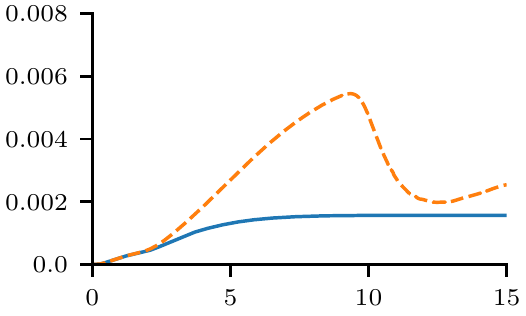}
      \caption{$\nu = 10^{-4};\ \eta = 10^{-3}$}%
      \label{fig:v-4r-3_kinetic_energy_ps}
    \end{subfigure}
    \hfill
    \begin{subfigure}{0.33\linewidth}
      \centering
      \includegraphics[width=1.0\linewidth]{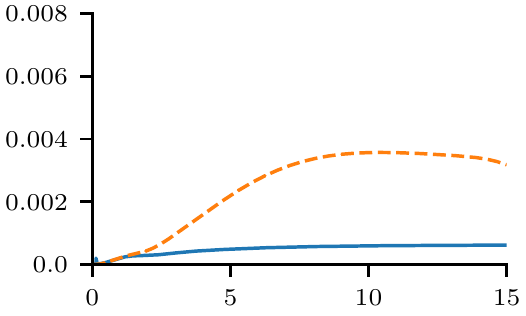}
      \caption{$\nu = 10^{-3};\ \eta = 10^{-3}$}%
      \label{fig:v-3r-3_kinetic_energy_ps}
    \end{subfigure}
    \hfill
    \begin{subfigure}{0.33\linewidth}
      \centering
      \includegraphics[width=1.0\linewidth]{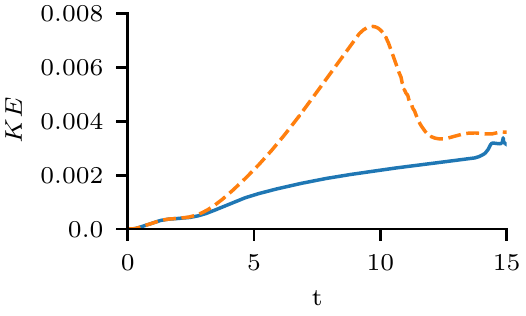}
      \caption{$\nu = 10^{-5};\ \eta = 10^{-4}$}%
      \label{fig:v-5r-4_kinetic_energy_ps}
    \end{subfigure}
    \hfill
    \begin{subfigure}{0.33\linewidth}
      \centering
      \includegraphics[width=1.0\linewidth]{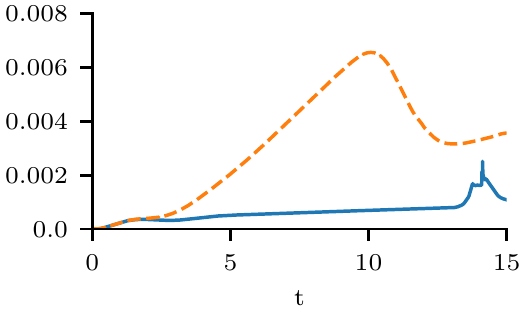}
      \caption{$\nu = 10^{-4};\ \eta = 10^{-4}$}%
      \label{fig:v-4r-4_kinetic_energy_ps}
    \end{subfigure}
    \hfill
    \begin{subfigure}{0.33\linewidth}
      \centering
      \includegraphics[width=1.0\linewidth]{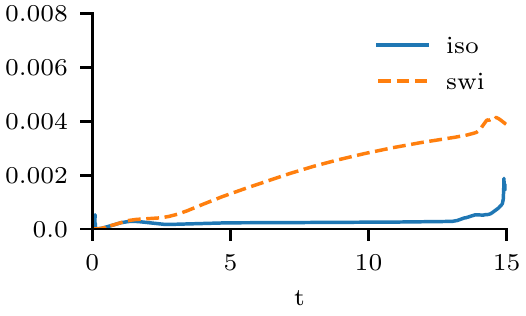}
      \caption{$\nu = 10^{-3};\ \eta = 10^{-4}$}%
      \label{fig:v-3r-4_kinetic_energy_ps}
    \end{subfigure}

  \mycaption{Kinetic energy as functions of time for each parameter choice and viscosity model.}{An increase in $\nu$ damps the energy released during the KHI in the switching cases and totally suppresses the KHI in most isotropic cases.}%
  \label{fig:param_study_kinetic_energies}
\end{figure*}

\subsubsection{Kinetic energy profiles}

Figure~\ref{fig:param_study_kinetic_energies} shows the kinetic energy as a function of time for all parameter choices and \rs{for}{} both viscosity models. The strongly KHI-unstable cases show a similar kinetic energy profile to that of the unstable typical case (figure~\ref{fig:v-4r-4_kinetic_energy}). In the switching cases, the peak kinetic energy is larger when $\eta=10^{-4}$ in two cases (figures~\ref{fig:v-5r-4_kinetic_energy_ps} and~\ref{fig:v-4r-4_kinetic_energy_ps}). This is a result of the reduced diffusion of the magnetic field resulting in a stronger vorticity layer. The isotropic cases in figure~\ref{fig:param_study_kinetic_energies} show an interesting trend in that the kinetic energy \rs{becomes stronger}{more} for smaller \rs{values of the viscosity}{} $\nu$ (as expected) or for larger \rs{values of the resistivity}{} $\eta$. In particular, the isotropic case where $\eta=10^{-3}$ and $\nu=10^{-5}$ (figure~\ref{fig:v-5r-3_kinetic_energy_ps}) is \rs{the}{} only isotropic case where the KHI is significantly unstable. In this case the kinetic energy profile shares a similar shape to the associated switching case, but the enhanced dissipation prevents the KHI from generating similar levels of kinetic energy. Instead, the profile is flatter and saturates at a later time.

Table~\ref{tab:stability} reveals three cases of interest which are explored through the kinetic energy profiles. The marginally unstable isotropic case, where $\eta=10^{-4}$ and $\nu=10^{-5}$, does show some growth but it is notably less than the fully unstable case (figure~\ref{fig:v-5r-4_kinetic_energy_ps}). Given that the current-vortex sheets in both these cases share similar strengths of vorticity, it is the combination of viscous dissipation of perturbations and enhanced magnetic shear in the lower $\eta$ case which acts to stabilise the sheet. This conclusion can similarly be drawn for the outlying switching case where $\eta=10^{-4}$ and $\nu=10^{-3}$ (figure~\ref{fig:v-3r-4_kinetic_energy_ps}). The remaining case of interest is where $\eta=10^{-3}$ and $\nu=10^{-4}$, the single case where the linear prediction disagrees with the observed stability (figure~\ref{fig:v-4r-3_kinetic_energy_ps}). In this case, the kinetic energy plateaus as the viscosity dissipates kinetic energy as it is generated by the instability.

\begin{figure*}[t]
    \centering
    \hfill
    \begin{subfigure}{0.33\linewidth}
  \includegraphics[width=1.0\linewidth]{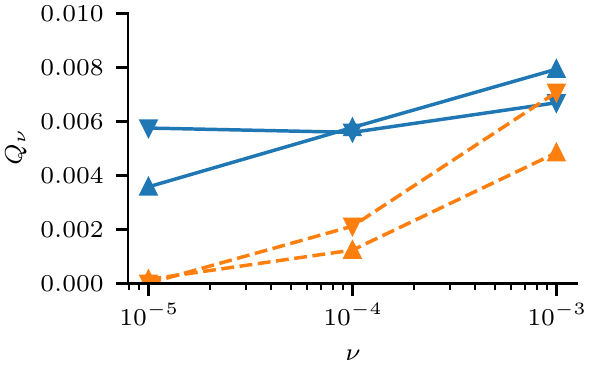}
      \caption{Viscous heating}%
      \label{fig:ps_visc_heating}
    \end{subfigure}
    \hfill
    \begin{subfigure}{0.33\linewidth}
  \includegraphics[width=1.0\linewidth]{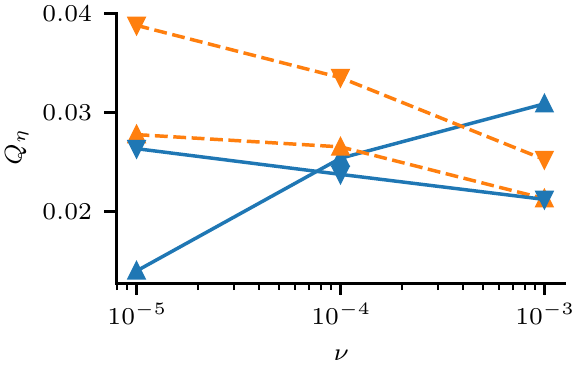}
      \caption{Ohmic heating}%
      \label{fig:ps_ohmic_heating}
    \end{subfigure}
    \hfill
    \begin{subfigure}{0.33\linewidth}
  \includegraphics[width=1.0\linewidth]{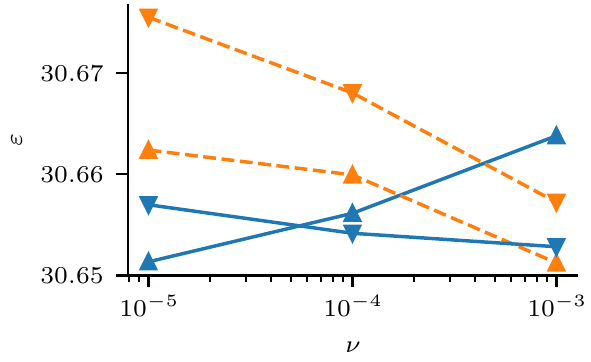}
      \caption{Internal energy}%
      \label{fig:ps_interal_energy}
    \end{subfigure}
  \mycaption{Internal energy, total viscous and total \rs{ohmic}{Ohmic} heating as functions of viscosity $\nu$ at $t=12.5$, before the onset of any null collapse, for all parameter choices.}{Isotropic (blue, solid) and switching viscosity (orange, dashed) are both shown. An upwards-facing triangle denotes the higher value of $\eta=10^{-3}$ and a downwards-facing triangle, $\eta=10^{-4}$. The anisotropic viscous heating can become significant for larger values of $\nu$ yet, when $\nu$ is smaller, the lack of viscous heating is compensated by enhanced \rs{ohmic}{Ohmic} heating.}%
  \label{fig:ps_heating}
\end{figure*}

\subsubsection{Heating profiles}

Figure~\ref{fig:ps_heating} presents the total heat generated by viscous and \rs{ohmic}{Ohmic} dissipation and the total internal energy at $t=13$, prior to any null collapse. Looking first at the viscous heating (figure~\ref{fig:ps_visc_heating}), $Q_{\nu}$ generally decreases with decreasing \rs{viscosity}{} $\nu$, as one may expect, with the exception of the isotropic cases where \rs{the resistvity}{} $\eta=10^{-4}$. Instead, in these cases the viscous heating shows little dependence on $\nu$. This reveals the complex, nonlinear relationship between viscous heating, the value of $\nu$ and the flows generated.

In the switching cases, generally an increase in $\nu$ increases viscous heating and decreases \rs{ohmic}{Ohmic} heating. The decrease in \rs{ohmic}{Ohmic} heating is due to two complementary effects. Firstly, viscosity generally slows flows and limits the compression of current sheets, consequently limiting \rs{ohmic}{Ohmic} heating, thus a larger $\nu$ produces less \rs{ohmic}{Ohmic} heating. Secondly, the nonlinear phase of the KHI enhances \rs{ohmic}{Ohmic} heating in the fan plane and, since the instability is more unstable for smaller $\nu$, \rs{ohmic}{Ohmic} heating increases with decreasing $\nu$. The overall effect is a decrease in internal energy with increasing $\nu$. This is also true for the $\eta=10^{-3}$ isotropic cases.

The \rs{ohmic}{Ohmic} heating profile similarly reveals complex behaviour in the isotropic cases (figure~\ref{fig:ps_ohmic_heating}. The stark difference in trends can be explained by considering the spatial distribution of \rs{ohmic}{Ohmic} heating which mirrors that of the current density. The two main current structures in a twisted null \rs{point}{} are the current-vortex sheet and the structure associated with the twisted spines (although the spine currents are two separate regions of current density, they contribute equally to the \rs{ohmic}{Ohmic} heating so are considered one here). These are the main sources of \rs{ohmic}{Ohmic} heating and the balance of contributions from each source, for different values of $\eta$ and $\nu$, is non-trivial and results in the observed difference in trends.

\begin{figure}[t]
  \begin{subfigure}{\linewidth}
      \centering
  \includegraphics[width=0.8\linewidth]{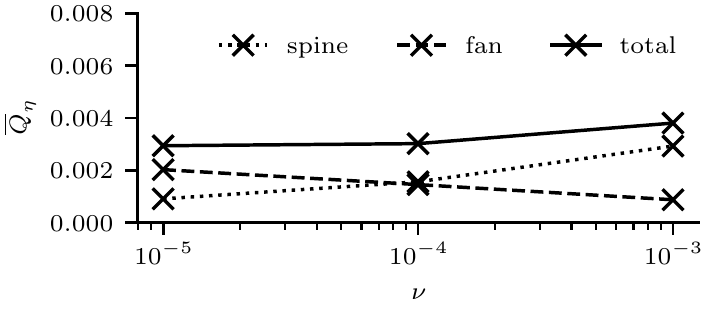}
      \caption{$\eta = 10^{-3}$}%
      \label{fig:balance_of_ohmic_heating-3}
    \end{subfigure}
    \hfill
    \begin{subfigure}{\linewidth}
      \centering
  \includegraphics[width=0.8\linewidth]{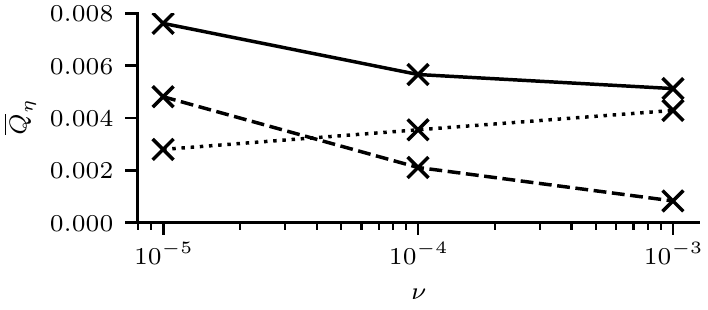}
      \caption{$\eta = 10^{-4}$}%
      \label{fig:balance_of_ohmic_heating-4}
    \end{subfigure}
  \mycaption{Ohmic heating contributions from separate current structures in the spine and fan.}{Shown are mean \rs{ohmic}{Ohmic} heating contributions from the spines (dotted) and current-vortex sheet (dashed) and their sum (solid) for $\eta=10^{-3}$ (a) and $\eta=10^{-4}$ (b) in only the isotropic cases at $t=10$. The value of $\eta$ dictates how rapidly the balance of contributions shifts from fan to spine with $\nu$, resulting in different trends in total \rs{ohmic}{Ohmic} heating.}%
  \label{fig:balance_of_ohmic_heating}
\end{figure}

Figure~\ref{fig:balance_of_ohmic_heating} reveals how the contributions from the current-vortex sheet and spines change with $\nu$, and the effect that has on the total \rs{ohmic}{Ohmic} heating. These measures are calculated as the mean of the \rs{ohmic}{Ohmic} heating in the $xy$-plane (representing the heating within the current-vortex sheet) and that in the $yz$-plane (representing the heating in the spines). These are not true measurements of the \rs{ohmic}{Ohmic} heating within each current structure, however they provide a useful proxy. 

For any value of $\eta$, the spine heating increases and the current-vortex heating decreases as $\nu$ increases. This is due to greater viscosity dissipating the initial Alfv\'en waves more effectively and reducing the magnetic shear in the current-vortex sheet while retaining magnetic shear in the spines. The difference in how rapidly the relative contributions change with $\nu$ gives rise to the difference in total \rs{ohmic}{Ohmic} heating trends found in figure~\ref{fig:ps_ohmic_heating}. When $\eta=10^{-4}$, the heating in the sheet decreases faster than the spine heating increases with $\nu$, resulting in a drop in total \rs{ohmic}{Ohmic} heating (figure~\ref{fig:balance_of_ohmic_heating-4}). The opposite is true when $\eta=10^{-3}$ (figure~\ref{fig:balance_of_ohmic_heating-3}).

\section{Discussion}
\label{sec:khi_discussion}

While the values of \rs{the resistivity}{} $\eta$ used in the simulations performed here are orders of magnitude greater than typical coronal estimates, the values of \rs{viscosity}{} $\nu$ are certainly within realistic bounds. It has been found that, when using a model of viscosity appropriate in the solar corona, i.e. the switching model, the KHI is unstable, regardless of parameter choice. This strongly suggests that, in the real corona, the KHI can be excited in current-vortex sheets similar to those studied here. %%What this investigation does not take into account are other possible null point configurations.

This \rs{paper}{chapter} details the investigation of the KHI and null point collapse around an axisymmetric, linear null point, an idealised model of a real null point (such as that observed in~\citet{massonNATUREFLARERIBBONS2009}). The impact of different null \rs{point}{} configurations such as those with asymmetry (e.g. those investigated \rs{by}{in}~\citet{thurgoodImplosiveCollapseMagnetic2018,pontinWhyAreFlare2016a}) is unclear. Similarly, the simplicity of the driver used here is unlikely to reflect the true nature of drivers in the real solar corona. The impact of driver complexity on spine-fan reconnection specifically has been investigated \rs{by}{}~\citet{wyperSpinefanReconnectionInfluence2012} \rs{who, however,}{however the drivers studied in~\citep{wyperSpinefanReconnectionInfluence2012}} focus on sheared drivers, as opposed to the torsional drivers employed here. It would be of interest to understand how different magnetic field configurations and forms of driver affect the formation and stability of the kind of current-vortex sheets studied here.

The simulations detailed here have been performed with a model of anisotropic viscosity which only captures the parallel component of viscosity. As discussed \rs{by}{in}~\citet{einaudiResistiveInstabilitiesFlowing1989}, perpendicular components can become significant in strong velocity shears (such as those found in the fan plane of a twisted null point) despite the small size of the associated transport coefficient. A similar set of experiments exploring the effect of perpendicular viscosity could provide useful insight, particularly in ascertaining if the growth of the tearing instability in the current-vortex sheet could be accelerated by perpendicular viscosity, as is found in the linear analysis performed by~\citet{einaudiResistiveInstabilitiesFlowing1989}. \dm{That being said, we expect that the inclusion of perpendicular effects will lead to differences that are quantitative (e.g. affecting the numerical size of the KHI growth rate) rather than qualitative. This is because the perpendicular components of viscosity act in localised regions   \citep[e.g.][]{mactaggartBraginskiiMagnetohydrodynamicsArbitrary2017}  and not everywhere in the domain like isotropic viscosity. Thus, we expect the effects of perpendicular viscosity to play a role more similar to parallel viscosity than isotropic viscosity.}{} 

An important finding of this investigation is the spontaneous collapse of the null point without shearing drivers (as in~\citet{pontinCurrentSheetFormation2007}) or prescribed current density perturbations (as in~\citet{thurgoodImplosiveCollapseMagnetic2018}). The formation of the current sheet at the null point (which facilitates spine-fan reconnection and null \rs{point}{} collapse) is primarily driven by a shearing motion at the null point, itself a result of oppositely directed streams of plasma flowing along the spine towards the null \rs{point}{}. These flows rely on pressure gradients which are generated by \rs{ohmic}{Ohmic} heating in the twisted spine. It may be that the pressure generated under ideal conditions (or using realistically small coronal resistivity) is not enough to drive the null-directed flows and, thus, not enough to collapse the null \rs{point}{}. Further investigation of this form of collapse within a twisted null point is required to ascertain if such a collapse is possible in the real corona.

The effect of the form of viscosity on the collapse of the null \rs{point}{} is also explored in the two high-resolution simulations (of section~\ref{sec:null_point_khi_single_case}) and it is found that in the switching case, where the KHI is unstable, the null \rs{point}{} collapses notably earlier than in the isotropic case, where the KHI is stable. It is unclear if the early null collapse is a consequence of the KHI or the use of the switching model. From the results of the isotropic case, the null \rs{point}{} collapse appears to be ultimately caused by slight asymmetries in the spine-aligned flows, so one may conjecture that, in the switching case, the KHI introduces its own asymmetries which cause the early collapse of the null \rs{point}{}. A higher resolution version of the unstable, isotropic case (where $\nu = 10^{-5}$ and $\eta=10^{-3}$) would provide clarity.

It is unclear how the unravelling of the null \rs{point}{} as it collapses affects the ability of the null \rs{point}{} to undergo the kind of oscillatory reconnection found \rs{by}{in}~\citet{thurgoodThreedimensionalOscillatoryMagnetic2017}. One observed phase in the oscillatory process is the generation of back-pressure which halts and reverses the spine-fan reconnection process. The simulations performed here do not run for a long enough time to investigate the generation of back-pressure. It may be that a collapsing twisted null \rs{point}{} is unable to produce the required back-pressure if it unravels during its initial collapse. Running the high-resolution simulations reported here for a longer time would reveal if the particular setup studied here can generate oscillatory spine-fan reconnection. Alternatively, using a pre-twisted null point as an initial condition with the perturbation used to collapse the null \rs{point}{} found in~\citet{thurgoodThreedimensionalOscillatoryMagnetic2017} would provide a similar experiment.

\section{Conclusions}
\label{sec:khi_conclusions}

In this paper two models of viscosity have been applied to a magnetic null point which has been dynamically twisted at its footpoints in such a way that a current-vortex sheet forms in the fan plane. This sheet has the potential to become unstable to the KHI. It was found that increased viscous dissipation, particularly in the form of isotropic viscosity, has a stabilising effect on the sheet, to the point of complete suppression of the instability. This is primarily due to viscosity thickening the sheet and increasing its stability. The presence of the instability enhances reconnection and viscous heating within the sheet.

After some time, the null \rs{point}{} spontaneously collapses due to an imbalance in spine-directed, pressure-driven flows. This was found to occur sooner when the KHI is present. The general development of the collapse and associated spine-fan reconnection is similar to that of previous work with the exception that the twist in the spine unravels during the collapse.

The investigation of the stability of the current-vortex sheet was extended with a parameter study over an order of magnitude difference in resistivity and two \rs{orders of magnitude}{} in viscosity. The results show that the KHI is mostly unstable when using anisotropic viscosity and mostly stable when using isotropic viscosity.

\dm{The general qualitative behaviour that we have found in this study matches well with our previous study on the effect of anisotropic viscosity on the kink instability in a flux tube. \citep{quinnEffectAnisotropicViscosity2020a}. The more localised influence of anisotropic viscosity compared to isotropic viscosity allows for the creation of much smaller length scales, both in the magnetic and velocity fields. The weaker effect of direct damping, by anisotropic viscosity compared to isotropic viscosity, means that the shorter length scales also coincide, in general, with higher magnitudes, thus enhancing the possibility of instability. Even with different dynamical drivers drivers (the null point is twisted in this study whereas the flux tube in \cite{quinnEffectAnisotropicViscosity2020a} is initially unstable and allowed to relax)  the cases with anisotropic viscosity exhibit the fast development of instabilities and reconnection in the way described above. Although anisotropic viscosity does not affect the magnetic field directly, it does have a significant indirect influence through nonlinear interaction. Therefore, in the study of nonlinear dynamics in the solar corona, taking account of anisotropic viscosity is important as it can result in the development of behaviour which would be absent if only isotropic viscosity were considered.}{}

\begin{acknowledgements}
Results were obtained using the ARCHIE-WeSt High Performance Computer
(\url{www.archie-west.ac.uk}) based at the University of
Strathclyde. JQ was funded via an EPSRC studentship: EPSRC DTG EP/N509668/1.
\end{acknowledgements}

\bibliographystyle{aa}
\bibliography{paper}

\begin{thebibliography}{40}
\expandafter\ifx\csname natexlab\endcsname\relax\def\natexlab#1{#1}\fi

\bibitem[{Antiochos {et~al.}(1999)Antiochos, DeVore, \&
  Klimchuk}]{AntiochosCME199}
Antiochos, S.~K., DeVore, C.~R., \& Klimchuk, J.~A. 1999, ApJ, 510, 485

\bibitem[{Arber {et~al.}(2001)Arber, Longbottom, Gerrard, \&
  Milne}]{arberStaggeredGridLagrangian2001}
Arber, T., Longbottom, A., Gerrard, C., \& Milne, A. 2001, J Comp Phys, 171,
  151

\bibitem[{Barnes(2007)}]{barnesRelationshipCoronalMagnetic2007}
Barnes, G. 2007, ApJ, 670, L53

\bibitem[{Bennett {et~al.}(2020)Bennett, {TonyArber}, Quinn, \&
  {csbrady-warwick}}]{keith_bennett_2020_4155546}
Bennett, K., {TonyArber}, Quinn, J.~J., \& {csbrady-warwick}. 2020,
  {{JamieJQuinn}}/Lare3d: {{Anisotropic}} Viscosity Feature Release for Thesis,
  Zenodo

\bibitem[{Braginskii(1965)}]{braginskiiTransportProcessesPlasma1965}
Braginskii, S.~I. 1965, Rev Plasma Phys, 1, 205

\bibitem[{Chandrasekhar(1961)}]{chandrasekhar1961}
Chandrasekhar, S. 1961, Hydrodynamic and {{Hydromagnetic Stability}} ({Oxford}:
  {Clarendon Press})

\bibitem[{Edwards \& Parnell(2015)}]{edwardsNullPointDistribution2015}
Edwards, S.~J. \& Parnell, C.~E. 2015, Sol Phys, 290, 2055

\bibitem[{Einaudi \& Rubini(1986)}]{einaudiResistiveInstabilitiesFlowing1986}
Einaudi, G. \& Rubini, F. 1986, The Physics of Fluids, 29, 2563

\bibitem[{Einaudi \& Rubini(1989)}]{einaudiResistiveInstabilitiesFlowing1989}
Einaudi, G. \& Rubini, F. 1989, Physics of Fluids B: Plasma Physics, 1, 2224

\bibitem[{Faganello \& Califano(2017)}]{faganelloMagnetizedKelvinHelmholtz2017}
Faganello, M. \& Califano, F. 2017, Journal of Plasma Physics, 83

\bibitem[{Foullon {et~al.}(2011)Foullon, Verwichte, Nakariakov, Nykyri, \&
  Farrugia}]{foullonMAGNETICKELVINHELMHOLTZINSTABILITY2011}
Foullon, C., Verwichte, E., Nakariakov, V.~M., Nykyri, K., \& Farrugia, C.~J.
  2011, ApJL, 729, L8

\bibitem[{Galsgaard(2003)}]{galsgaardNumericalExperimentsWave2003}
Galsgaard, K. 2003, J Geophys Res, 108

\bibitem[{Galsgaard \& Pontin(2011)}]{galsgaardSteadyStateReconnection2011}
Galsgaard, K. \& Pontin, D.~I. 2011, A\&A, 529, A20

\bibitem[{Hollweg(1986)}]{hollwegViscosityChewGoldbergerLowEquations1986a}
Hollweg, J.~V. 1986, ApJ, 306, 730

\bibitem[{Howson {et~al.}(2017)Howson, Moortel, \&
  Antolin}]{howsonEffectsResistivityViscosity2017}
Howson, T.~A., Moortel, I.~D., \& Antolin, P. 2017, A\&A, 602, A74

\bibitem[{Kowal {et~al.}(2020)Kowal, {Falceta-Gon{\c c}alves}, Lazarian, \&
  Vishniac}]{kowalKelvinHelmholtzTearingInstability2020}
Kowal, G., {Falceta-Gon{\c c}alves}, D.~A., Lazarian, A., \& Vishniac, E.~T.
  2020, ApJ, 892, 50

\bibitem[{Maclean {et~al.}(2005)Maclean, Beveridge, Longcope, Brown, \&
  Priest}]{macleanTopologicalAnalysisMagnetic2005}
Maclean, R., Beveridge, C., Longcope, D., Brown, D., \& Priest, E. 2005,
  Proceedings of the Royal Society A: Mathematical, Physical and Engineering
  Sciences, 461, 2099

\bibitem[{MacTaggart {et~al.}(2017)MacTaggart, Vergori, \&
  Quinn}]{mactaggartBraginskiiMagnetohydrodynamicsArbitrary2017}
MacTaggart, D., Vergori, L., \& Quinn, J. 2017, J Fluid Mech, 826, 615

\bibitem[{Masson {et~al.}(2009)Masson, Pariat, Aulanier, \&
  Schrijver}]{massonNATUREFLARERIBBONS2009}
Masson, S., Pariat, E., Aulanier, G., \& Schrijver, C.~J. 2009, ApJ, 700, 559

\bibitem[{McLaughlin {et~al.}(2012)McLaughlin, Thurgood, \&
  MacTaggart}]{McLauglinOscRec2012}
McLaughlin, J.~A., Thurgood, J.~O., \& MacTaggart, D. 2012, A\&A, 548, A98

\bibitem[{Min {et~al.}(1997)Min, Kim, \&
  Lee}]{minEffectsMagneticReconnection1997}
Min, K.~W., Kim, T., \& Lee, H. 1997, Planetary and Space Science, 45, 495

\bibitem[{Miyama {et~al.}(2012)Miyama, Tomisaka, \&
  Hanawa}]{miyamaNumericalAstrophysicsProceedings2012}
Miyama, S.~M., Tomisaka, K., \& Hanawa, T. 2012, Numerical {{Astrophysics}}:
  {{Proceedings}} of the {{International Conference}} on {{Numerical
  Astrophysics}} 1998 ({{NAP98}}), Held at the {{National Olympic Memorial
  Youth Center}}, {{Tokyo}}, {{Japan}}, {{March}} 10\textendash 13, 1998
  ({Springer Science \& Business Media})

\bibitem[{{Moreno-Insertis} \&
  Galsgaard(2013)}]{moreno-insertisPLASMAJETSERUPTIONS2013}
{Moreno-Insertis}, F. \& Galsgaard, K. 2013, ApJ, 771, 20

\bibitem[{Pontin {et~al.}(2016)Pontin, Galsgaard, \&
  D{\'e}moulin}]{pontinWhyAreFlare2016a}
Pontin, D., Galsgaard, K., \& D{\'e}moulin, P. 2016, Sol Phys, 291, 1739

\bibitem[{Pontin {et~al.}(2007)Pontin, Bhattacharjee, \&
  Galsgaard}]{pontinCurrentSheetFormation2007}
Pontin, D.~I., Bhattacharjee, A., \& Galsgaard, K. 2007, Physics of Plasmas,
  14, 052106

\bibitem[{Priest {et~al.}(2003)Priest, Hornig, \&
  Pontin}]{priestNatureThreedimensionalMagnetic2003}
Priest, E.~R., Hornig, G., \& Pontin, D.~I. 2003, J Geophys Res, 108, 1285

\bibitem[{Quinn {et~al.}(2020)Quinn, MacTaggart, \&
  Simitev}]{quinnEffectAnisotropicViscosity2020a}
Quinn, J., MacTaggart, D., \& Simitev, R.~D. 2020, Communications in Nonlinear
  Science and Numerical Simulation, 83, 105131

\bibitem[{Quinn(2021)}]{quinnJamieJQuinnKhiNull2021}
Quinn, J.~J. 2021, {{JamieJQuinn}}/Khi\_null\_point\_code: {{Initial}}
  Submission to {{A}}\&{{A}}, Zenodo

\bibitem[{Roediger {et~al.}(2013)Roediger, Kraft, Nulsen, Churazov, Forman,
  Br{\"u}ggen, \&
  Kokotanekova}]{roedigerViscousKelvinHelmholtzInstabilities2013a}
Roediger, E., Kraft, R.~P., Nulsen, P., {et~al.} 2013, Monthly Notices of the
  Royal Astronomical Society, 436, 1721

\bibitem[{Ryu {et~al.}(2000)Ryu, Jones, \&
  Frank}]{ryuMagnetohydrodynamicKelvinHelmholtzInstability2000}
Ryu, D., Jones, T.~W., \& Frank, A. 2000, ApJ, 545, 475

\bibitem[{Schindler {et~al.}(1988)Schindler, Hesse, \&
  Birn}]{schindlerGeneralMagneticReconnection1988}
Schindler, K., Hesse, M., \& Birn, J. 1988, J Geophys Res, 93, 5547

\bibitem[{Sun {et~al.}(2013)Sun, Hoeksema, Liu, Aulanier, Su, Hannah, \&
  Hock}]{sunHOTSPINELOOPS2013}
Sun, X., Hoeksema, J.~T., Liu, Y., {et~al.} 2013, ApJ, 778, 139

\bibitem[{Thurgood {et~al.}(2017)Thurgood, Pontin, \&
  McLaughlin}]{thurgoodThreedimensionalOscillatoryMagnetic2017}
Thurgood, J.~O., Pontin, D.~I., \& McLaughlin, J.~A. 2017, The Astrophysical
  Journal, 844, 2

\bibitem[{Thurgood {et~al.}(2018)Thurgood, Pontin, \&
  McLaughlin}]{thurgoodImplosiveCollapseMagnetic2018}
Thurgood, J.~O., Pontin, D.~I., \& McLaughlin, J.~A. 2018, ApJ, 855, 50

\bibitem[{{TonyArber} {et~al.}(2020){TonyArber}, Bennett, Quinn, \&
  {csbrady-warwick}}]{keith_bennett_2020_4155646}
{TonyArber}, Bennett, K., Quinn, J.~J., \& {csbrady-warwick}. 2020,
  {{JamieJQuinn}}/Lare3d: {{KHI}} in Null Point Configuration Release for
  Thesis, Zenodo

\bibitem[{Wyper {et~al.}(2012)Wyper, Jain, \&
  Pontin}]{wyperSpinefanReconnectionInfluence2012}
Wyper, P.~F., Jain, R., \& Pontin, D.~I. 2012, A\&A, 545, A78

\bibitem[{Wyper \&
  Pontin(2013)}]{wyperKelvinHelmholtzInstabilityCurrentvortex2013}
Wyper, P.~F. \& Pontin, D.~I. 2013, Physics of Plasmas, 20, 032117

\bibitem[{Yang {et~al.}(2018)Yang, Xu, Lim, Kim, Cho, Kim, Chae, Cho, \&
  Ji}]{yangObservationKelvinHelmholtz2018}
Yang, H., Xu, Z., Lim, E.-K., {et~al.} 2018, ApJ, 857, 115

\bibitem[{Yang {et~al.}(2020)Yang, Zhang, Xu, Zhang, Zhong, \&
  Guo}]{yangImagingSpectralStudy2020}
Yang, S., Zhang, Q., Xu, Z., {et~al.} 2020, arXiv:2005.09613 [astro-ph]
  [\eprint[arXiv]{2005.09613}]

\bibitem[{Zou {et~al.}(2020)Zou, Jiang, Wei, Feng, Zuo, \&
  Wang}]{zouContinuousNullPointMagnetic2020}
Zou, P., Jiang, C., Wei, F., {et~al.} 2020, ApJ, 890, 10

\end{thebibliography}

\appendix
\section{Associated software}
\rs{\subsection{Anisotropic viscosity module}}{}

A custom version of Lare3d \rs{\citep{arberStaggeredGridLagrangian2001} has been developed where a new module for anisotropic viscosity has been included}{which implements the anisotropic viscosity module}. \rs{The new version}{} can be found at \url{https://github.com/jamiejquinn/Lare3d}, \rs{and also}{has been} archived at~\cite{keith_bennett_2020_4155546}, and should be simple to merge into another version of Lare3d for future research. The version of Lare3d used in the production of the results presented here, including initial conditions, boundary conditions, control parameters and the anisotropic viscosity module, can be found at~\cite{keith_bennett_2020_4155646}. The data analysis and instructions for reproducing all results found in this report may be also found at \url{https://github.com/jamiejquinn/khi_null_point_code} and has been archived at~\cite{quinnJamieJQuinnKhiNull2021}.

All simulations were performed on a single, multi-core machine with $40$ cores and $192$ GB of RAM, although this amount of RAM is much higher than was required; a conservative estimate of the memory used in the largest simulations is around $64$ GB. Most simulations completed in under $2$ days, although the longest running simulations (the highest-resolution cases shown here) completed in around $2$ weeks.

\subsection{Field line integrator}

As described in section~\ref{sec:reconn_rate}, the reconnection rate local to a single field line is given by the electric field parallel to the magnetic field, integrated along the field line. The global reconnection rate for a given region of magnetic diffusion is the maximum value of the local reconnection rate over all field lines threading the region. A field line integrator was developed specifically for this calculation and is detailed here.

Magnetic field lines lie tangential to the local magnetic field at every point $\vec{x}(s)$ along the line,
\begin{equation}
  \label{eq:field_line_equation}
  \frac{d\vec{x}(s)}{ds} = \vec{b}(\vec{x}(s)),
\end{equation}
where $s$ is a variable which tracks along a single field line and $\vec{b}$ is the unit vector in the direction of $\vec{B}$. This equation is discretised using a second-order Runge-Kutta scheme to iteratively calculate the discrete positions $\vec{x}_i$ along a field line passing through some seed position $\vec{x}_0$,
\begin{align}
  \label{eq:field_line_calculation}
  \vec{x}_{i+1} &= \vec{x}_i + h\vec{b}(\vec{x}'_i),\\
  \vec{x}'_i &= \vec{x}_i + \tfrac{h}{2}\vec{b}(\vec{x}_i)\rs{,}{}
\end{align}
where $h$ is a small step size. Since $\vec{b}$ is discretised, the value at an arbitrary location $\vec{x}_i$ is calculated using a linear approximation. The integration of a scalar variable $y$ is carried out along a field line given by a sequence of $N$ locations $\vec{x}_i$ using the midpoint rule,
\begin{equation}
  \label{eq:midpoint_rule}
  Y = \sum_{i=1}^{N} \frac{(y(\vec{x}_{i-1}) + y(\vec{x}_{i}))}{2},
\end{equation}
where $Y$ is the result of the integration. In practice, $N$ is not specified and the discretised field line contains the required number of points to thread from its seed location to the boundary of the domain.

While the linear interpolation, second-order Runge-Kutta and midpoint rule are all low order methods, testing higher-order methods showed little change in results but dramatically increased the runtime of the analysis. The lower-order methods used offer an acceptable compromise between speed and accuracy. The above algorithm is implemented in Python and can be found in the code directory linked in the Associated software section above, in \verb|shared/field_line_integrator.py| with examples of use in \verb|main/field_line_integrator.Rmd|. The integration of multiple field lines is an embarrassingly parallel problem and is parallelised in a straight-forward manner using a pool of threads supplied by the \verb|Pool| feature of the Python library \verb|multiprocessing|. Although the integrator is used solely to integrate the parallel electric field along magnetic field lines in this \rs{study}{thesis}, the tool can be easily applied to arbitrary vector and scalar fields.

%\begin{acknowledgements}
%Results were obtained using the ARCHIE-WeSt %High Performance Computer
%(\url{www.archie-west.ac.uk}) based at the %University of
%Strathclyde. JQ was funded via an EPSRC studentship: EPSRC DTG EP/N509668/1.
%\end{acknowledgements}

%\bibliographystyle{aa}
%\bibliography{paper}

\end{document}